\documentclass[aps,prl,twocolumn,superscriptaddress]{revtex4-1}
\pdfoutput=1
\usepackage{amsmath}
\usepackage{amssymb}
\usepackage{graphicx}
\usepackage[dvipsnames]{xcolor}
\usepackage{bm}% bold math
\usepackage[colorlinks,linkcolor=blue,citecolor=blue]{hyperref}

\begin{document}
\title{Topological quantum walks:
theory and experiments}
\author{Jizhou Wu}
\email{wujizhou@mail.ustc.edu.cn}
\affiliation{%
Shanghai Branch, National Laboratory for Physical Sciences at Microscale,
University of Science and Technology of China, Shanghai 201315, China%
}
\affiliation{%
CAS Center for Excellence and Synergetic Innovation Center in Quantum Information and Quantum Physics, University of Science and Technology of China,
Hefei, Anhui 230026, China%
}
\author{Wei-Wei Zhang}
\email{ww.zhang@sydney.edu.au}
\affiliation{Centre for Engineered Quantum Systems, School of Physics, The University of Sydney, Sydney, New South Wales 2006, Australia}
\author{Barry C.\ Sanders}
\email{bsanders@ustc.edu.cn}
\affiliation{%
Shanghai Branch, National Laboratory for Physical Sciences at Microscale,
University of Science and Technology of China, Shanghai 201315, China%
}
\affiliation{%
CAS Center for Excellence and Synergetic Innovation Center in Quantum Information and Quantum Physics, University of Science and Technology of China, Hefei, Anhui 230026, China%
}
\affiliation{%
Institute for Quantum Science and Technology, University of Calgary, Alberta T2N 1N4, Canada%
}
\date{\today}
\maketitle
Inspired by recent breakthroughs with topological quantum materials,
which pave the way 
to novel, high-efficiency, low-energy
magnetoelectric devices~\cite{Li2014NN,Ando2014NL,DC2018NM}
and fault-tolerant quantum information processing~\cite{Kitaev2003APb},
inter alia,
topological quantum walks have emerged as an exciting topic in its own right,
especially due to the theoretical and experimental simplifications this approach offers~\cite{Kitagawa2010PRA,Kitagawa2010PRB,Kitagawa2012NC,Asboth2012PRB,Rudner2013PRX,Zhang2017PRL,Xiao2017NP,Chen2018PRL,Sun2018PRL,Sajid2019PRB}.
Motivated by impressive progress in topological quantum walks, 
we provide a perspective on theoretical studies and experimental investigations of topological quantum walks
focusing on current explorations of topological properties arising for single-walker quantum walks.

Quantum walk history traces back to 1993~\cite{Kempe2003CP}
when
Yakir Aharonov, Luis Davidovich and Nicim Zagury proposed the notion of the ``quantum random walk''
with this nomenclature indicating a quantum version 
of the ubiquitous random walk
in classical physics~\cite{Aharonov1993PRA}.
Eight years later, 
computer scientist Dorit Aharonov,
who coincidentally is Yakir Aharonov's niece,
and her colleagues
Andris Ambainis, Julia Kempe and Umesh Vazirani,
introduced a quantum walk on a general regular graph $\mathcal{G}$~\cite{Aharonov2001STOC}.
At the same conference,
quantum walks on one-dimensional lattices with Hadamard coin operators
(coherently flipping a coin to a superposition of `head' and `tails')
was introduced to compare with the case 
of the symmetric random walk~\cite{Ambainis2001STOC}.
One year later,
one-dimensional quantum walks were generalised to two- and higher-dimensional quantum walks~\cite{Mackay2002JPA}.

Quantum walks have become germane to quantum computation~\cite{Childs2003STOC,Ambainis2007SJC,Magniez2007SJC,Farhi2008TC,Douglas2008JPA,Childs2009PRL}
and quantum simulation~\cite{Godoy1992JCP,Mulken2011PR,DiMolfetta2013PRA,Zhang2016NJP}
and single-walker versions are amenable
to experimental implementations
including ion traps~\cite{Schmitz2009PRL,Zahringer2010PRL},
superconducting systems~\cite{Flurin2017PRX,Yan2019Science},
nuclear magnetic resonance~\cite{Du2003PRA,Ryan2005PRA},
optical lattice~\cite{Karski2009Science,Dadras2018PRL},
and both free-space linear optics~\cite{Do2005JOSAB,Cardano2015SA}
and on photonic chips~\cite{Perets2008PRL,Tang2018SA}.
Quantum walks are typically classified into discrete- 
and continuous-time quantum walks
with the main difference being
whether free evolution is interrupted by quantum coin `flips' followed by coin-dependent translations
or whether evolution is continuous
involving entangling between the walker and internal, or coin,
degrees of freedom~\cite{Venegas-Andraca2012QIP}.
For a discrete-time quantum walk on $\mathcal{G}$ of degree~$m$, 
the Hilbert-space dimension of the coin is $m$~\cite{Ambainis2001STOC,Aharonov2001STOC}.
As coin and translation operators act repeatedly on the walker space,
the the discrete-time quantum-walk
Hamiltonian is periodic in the time domain,
which indicates that the discrete-time quantum walk is a Floquet system.
Continuous-time quantum evolution on $\mathcal{G}$ is governed by a static Hamiltonian, 
which is the adjacency matrix for $\mathcal{G}$~\cite{Farhi1998PRA}. 
Any Hamiltonian describing a lattice model can equivalently be used to drive a continuous-time quantum walk on corresponding~$\mathcal{G}$
such as the Su-Schrieffer-Heeger model~\cite{Su1979PRL}
or spin-orbit coupling Hamiltonian~\cite{Sun2018PRL}. 
Discrete- and continuous-time quantum walks
are approximately equivalent under a Trotter approximation in the continuous-time limit~\cite{Schmitz2016PLA}.

Now we connect the quantum walk to topology,
which concerns global properties
that are invariant under continuous deformation
such as the geometric phase~\cite{Berry1984PRSLA} and quantum Hall effect~\cite{Klitzing1980PRL}.
In quantum walks, 
topology is characterised
by the integer-labelled homotopy group
corresponding to the mapping
from the Brillouin zone
to the space of unitary evolution operators or their Hamiltonian generators~\cite{Kitagawa2010PRB}.
The homotopy group label
is the homotopic, or topological, invariant,
and examples of such labels include the winding number~\cite{Kitagawa2010PRB}, Chern number~\cite{Chern1946AM} and Chern-Simons number~\cite{Chern1974AM}.
Topological properties can be investigated in bulk dynamics for a homogeneous quantum walk with fixed topological invariants~\cite{Cardano2016NC,Cardano2017NC,Zhang2017PRL,Sun2018PRL,Wang2019PRL}.
Alternatively, topological properties can be studied for boundary effects such as edge states between two bulk regions for an inhomogeneous quantum walk~\cite{Kitagawa2012NC,Groh2016PRA,Mugel2016PRA},
which could have different Hamiltonian parameters for differing bulk regions but the same winding number~\cite{Kitagawa2010PRB,Kitagawa2012NC}.

Topological properties of quantum walks
were first studied for the discrete-time quantum walks, called split-step quantum walks, 
which split the evolution into
alternating coin-dependent shifts
and coherent coin flips~\cite{Kitagawa2010PRA}.
Experimental investigations of topological split-step quantum walks
are photonic in nature~\cite{Kitagawa2012NC,Cardano2016NC,Cardano2017NC,Xiao2017NP,Chen2018PRL, Nitsche2019NJP}.
In photonic systems, the walker's position can be encoded into path~\cite{Kitagawa2012NC},
time-bin~\cite{Chen2018PRL,Nitsche2019NJP} or orbital-angular momentum~\cite{Cardano2016NC,Cardano2017NC} degrees of freedom.
This position can be converted to the Brillouin zone via a Fourier transformation,
which is convenient for studying topological properties in the bulk case.
The coin is encoded into an internal
photonic degree of freedom,
such as polarisation
or spin-angular momentum~\cite{Cardano2016NC,Cardano2017NC}.

The first observation of topological phenomena was achieved with 
a seven-step photonic one-dimensional split-step quantum walk
with the walker's position and coin encoded into diffraction-based path and polarisation degrees of freedom, respectively, for heralded single photons
generated by spontaneous parametric down conversion
~\cite{Kitagawa2012NC}.
By fixing one coin-rotation parameter
and making the other coin-rotation parameter position-dependent, 
a topological boundary is created
such that the topology characterised by winding numbers at the left and right sides of the boundary differ. 
Initialising the walker at the lattice boundary
ensures prominent localisation of the walker as an edge state.
The empirical signature of walker localisation is realised 
by repeatedly sampling the time-dependent position marginal distribution at each time
with the coin degree of freedom traced out.
To measure the sample at each time step,
the state needs to be re-initialised at the boundary and evolve coherently until measuring position at the specified time.
This procedure can be repeated for each desired evolution time.

Equivalence between a single step of a one-dimensional split-step quantum walk 
and a concatenating two steps of an ordinary one-dimensional quantum walk, with different
coin operators at each of the pair of steps,
ushered in experimental studies of topological quantum walks~\cite{Zhang2017PRA,Nitsche2019NJP}. 
The topology of concatenated ordinary quantum walks is enabled
by choosing different coin-operator parameters for each ordinary quantum walk operator in a single step.
Inhomogeneity arising by choosing different coin-operator parameters at different lattice sites enables the creation of topological boundaries.
Topological signatures are evident through measurements 
that yield eigenvalues and the approximate eigenstates
for concatenated ordinary quantum walks realised 
through the well established time-multiplexing architecture~\cite{Nitsche2019NJP}.

If the discrete-time quantum-walk Hamiltonian has chiral~\cite{Kitagawa2012NC} or sublattice symmetry~\cite{Asboth2012PRB}
(which could be equivalent~\cite{Yao2017PRB}),
each eigenstate with energy~$E$
has a counterpart with opposite energy~$-E$
except for $E\in\{0,\pm\pi\}$.
Between two bulk regions with the same winding number but different Hamiltonian parameters,
a pair of edge states,
localised to this boundary,
appear and thereby opens the door to exploring topological Floquet systems.
Anti-commutation between the time-dependent Hamiltonian of a discrete-time quantum walk 
and the chiral or sublattice symmetry operator guarantees topological band structures, which are evidenced by the topological edge states around zero and $\pi$ energy in the spectrum~\cite{Asboth2012PRB,Yao2017PRB}.
In photonic experiments,
these two edge states appear
as period-two oscillations of the walker's localised position
at the boundary~\cite{Kitagawa2012NC}. 

Existence of pair edge states indicates that a single winding number is insufficient
for topological classification of one-dimensional discrete-time quantum walks
with chiral~(sublattice) symmetry.
A pair of topological invariants extracted from eigenvalues of the chiral (sublattice) symmetry operator~\cite{Kitagawa2012NC}, 
which can also be directly measured from the scattering matrix in the view of scattering theory of quantum walks~\cite{Tarasinski2014PRA,Barkhofen2017PRA},
or a pair of winding numbers with extra topological information of time-shifted unitary evolution operators~\cite{Asboth2013PRB,Asboth2014PRB,Obuse2015PRB},
have been used for topological classification. 
More generally,
the topology of quantum walks on an infinite one-dimensional lattice can be completely 
classified with respect to the tenfold discrete symmetry groups constructed from all combinations of chiral,
time-reversal and particle-hole symmetries~\cite{Cedzich2016JPA,Cedzich2018AHP,Cedzich2018Quantum}. 
The Schur approach,
which transcends the standard Fourier transformation,
is utilised for the complete classification of non-translation invariant phases~\cite{Cedzich2019arXiv}.

The Chern number does not suffice for studying topology of two-dimensional discrete-time quantum walks,
whereas the Rudner-Lindner-Berg-Levin (RLBL) invariant can serve as the homotopy invariant, 
which distinguishes between static and driven two-dimensional systems~\cite{Rudner2013PRX}. 
By considering
both time and momentum domains,
the RLBL invariant provides additional information
for the existence
of the ``anomalous'' spectra that arise in driven systems.
The RLBL invariant is able to identify
phenomena
that cannot be characterised by the Chern number,
such as the appearance of robust chiral edge states in two-dimensional driven systems 
even though the Chern numbers of all the bulk Floquet bands are zero~\cite{Asboth2015PRA}.
By constructing an Abelian gauge field---the analogue of a magnetic flux threading a two-dimensional electron gas, 
the two-dimensional discrete-time magnetic quantum walk is proposed 
to be able to simulate anomalous Floquet Chern topological insulators, 
where the RLBL invariant is also necessary to 
fully characterise the corresponding anomalous Floquet topological phases 
and to compute the number of topologically protected edge modes~\cite{Sajid2019PRB}.

Bulk-edge correspondence manifests as robust unidirectional walker motion at the edge between two bulk regions
with different RLBL invariants
in two-dimensional discrete-time quantum walks~\cite{Asboth2015PRA}.
Two photonic experiments
first manifested such robust chiral edge states in two-dimensional split-step quantum walks~\cite{Chen2018PRL,Wang2018PRLb}.
One evolves up to 4 steps in the angular-orbital momentum space with the polarisation as a coin~\cite{Wang2018PRLb}. 
The other evolves up to 25 steps with time-bin and polarisation degrees of freedom as the walker's position and coin state, respectively~\cite{Chen2018PRL}. 
The walker's evolution in the time-bin space is implemented by two polarising beam-splitter loops,
one large and one small.
The large loop splits and delays each pulse into two pulses, 
which corresponds to the walker moving forward vs backwards over one distance unit of one coordinate axis in the synthetic walker's position space.
The small loop drives the walker's evolution 
along another coordinate axis in the two-dimensional walker's synthetic
position space.

An alternative to studying topological properties of inhomogeneous discrete-time quantum walks by initialising at an edge
is instead to study topological properties for
edge-less homogeneous discrete-time quantum walks
initialised anywhere.
For this case,
we can directly measure topological invariants by the reconstruction of eigenstates in the discrete-time quantum walks~\cite{Xu2018PRL} or by the interference of the walker with the accumulated Berry phase in the Bloch oscillating quantum walks~\cite{Ramasesh2017PRL,Flurin2017PRX},
or
study signatures of topological effects in the walker's dynamics,
with these signatures revealing effects
such as a transition,
between topological invariants 
achieved via adjusting a unitary-evolution parameter.
A topological transition can be 
revealed through the walker's spreading rate as a function of Hamiltonian parameters and evolution time~\cite{Cardano2016NC}.
Another signature is the separation of the mean position for the walker with a coin in one eigenstate of the chiral operator vs the mean position for the walker with the other eigenstate of the chiral operator,
known as mean chiral displacement,
for a quantum walk with chiral symmetry~\cite{Cardano2017NC}.
The experiments to study these two topological transition signals were conducted for photonic systems
with orbital-angular momentum as the walker's synthetic position space
and spin-angular momentum as the coin space.

Analogous to topological studies of discrete-time quantum walks,
edge-state~\cite{Blanco-Redondo2016PRL}
and topological-transition~\cite{Zhang2017PRL,Wang2019PRL}
signatures manifest for continuous-time topological quantum walks as well.
Such continuous-time topological quantum walks have been explored 
experimentally in silicon waveguides~\cite{Blanco-Redondo2016PRL} and in photonic chips~\cite{Wang2019PRL}.
The band-inversion ring is another
topological pattern 
that can be directly extracted from 
measuring the continuous-time walker's momentum~\cite{Zhang2018SB},
with the walker being a cold $^{87}$Rb atom in an optical lattice.
The quantum walk is initiated by a quench,
meaning a sudden change of a Hamiltonian parameter~\cite{Sun2018PRL}.
The topological signature is 
revealed in the walker's 
momentum-space spin-polarisation distribution 
experiments.

In the optical-lattice cold-atom experiments,
synthetic spin is simulated by two magnetic sub-levels
that are coupled together
via two orthogonal Raman laser beams, 
and the walker's position is encoded in true position space~\cite{Sun2018PRL}. 
An atom cloud initially prepared in the topologically trivial state 
evolves under a topologically non-trivial Hamiltonian after the quench. 
Appearance of a ring pattern in momentum space during evolution signals a non-trivial topology of the post-quench Hamiltonian,
and momentum-component oscillations
reveal band structure. 
Existence of a ``kink'' for the walker's expansion rate in lattice space reveals a topological transition of the system~\cite{Zhang2017PRL}. 

Thus far,
we have considered standard quantum mechanics,
but topological quantum walks can be studied in the alternative parity-time-symmetric
(PT-symmetric)
quantum mechanics,
which is an alternative form of quantum mechanics based on modifying the inner product according to the principle that the Hamiltonian can be non-Hermitian provided that its spectrum is real~\cite{Bender1998PRL}.
Topologically protected edge states exist as well in PT-symmetric topological quantum walks,
shown by driving a PT-symmetric nonunitary walker evolution~\cite{Xiao2017NP}.
Parity-time symmetry is enforced
in experimental one-dimensional discrete-time
parity-time-symmetric photonic quantum walks 
by introducing alternating lossy steps that are interpreted as alternating balanced noiseless loss and gain,
through modifying the norm by postselection~\cite{Mochizuki2016PRA,Xiao2017NP}.
A topological quantum walk has been demonstrated for this system 
up to six steps.
This six-step quantum walk reveals topological edge states at the boundary between two bulks with distinct winding number pairs
in the inhomogeneous case~\cite{Xiao2017NP}
and the topologically protected skyrmion in the time-momentum space for the homogeneous case with quench~\cite{Wang2019NC}.

Quantum walks are a powerful concept and tool for studying 
topological effects arising from profound considerations that arise in studies of quantum materials and their experimental realisations.
In the future, more general topological quantum walks,
involving for example high-dimensional coins and multiple interacting walkers will teach much about more exotic topological effects.
Another exciting direction concerns
the use of machine learning to identify topological phases~\cite{Ming2018arXiv} and topological phase transitions~\cite{Rem2019NP}
in quantum walks.
\begin{acknowledgments}
\emph{Acknowledgments} B.\ C.\ S.\ and J.\ W.\ are supported by the National Natural Science Foundation of China~(NSFC) with grant No.\ 11675164. W.\ Z.\ is supported by the Australian Research Council~(ARC) via the Centre of Excellence in Engineered Quantum Systems~(EQuS) project number CE110001013, and USyd-SJTU Partnership Collaboration Awards.
\end{acknowledgments}

\bibliographystyle{apsrev4-1}
\bibliography{topoqw}

\begin{thebibliography}{77}%
\makeatletter
\providecommand \@ifxundefined [1]{%
 \@ifx{#1\undefined}
}%
\providecommand \@ifnum [1]{%
 \ifnum #1\expandafter \@firstoftwo
 \else \expandafter \@secondoftwo
 \fi
}%
\providecommand \@ifx [1]{%
 \ifx #1\expandafter \@firstoftwo
 \else \expandafter \@secondoftwo
 \fi
}%
\providecommand \natexlab [1]{#1}%
\providecommand \enquote  [1]{``#1''}%
\providecommand \bibnamefont  [1]{#1}%
\providecommand \bibfnamefont [1]{#1}%
\providecommand \citenamefont [1]{#1}%
\providecommand \href@noop [0]{\@secondoftwo}%
\providecommand \href [0]{\begingroup \@sanitize@url \@href}%
\providecommand \@href[1]{\@@startlink{#1}\@@href}%
\providecommand \@@href[1]{\endgroup#1\@@endlink}%
\providecommand \@sanitize@url [0]{\catcode `\\12\catcode `\$12\catcode
  `\&12\catcode `\#12\catcode `\^12\catcode `\_12\catcode `\%12\relax}%
\providecommand \@@startlink[1]{}%
\providecommand \@@endlink[0]{}%
\providecommand \url  [0]{\begingroup\@sanitize@url \@url }%
\providecommand \@url [1]{\endgroup\@href {#1}{\urlprefix }}%
\providecommand \urlprefix  [0]{URL }%
\providecommand \Eprint [0]{\href }%
\providecommand \doibase [0]{http://dx.doi.org/}%
\providecommand \selectlanguage [0]{\@gobble}%
\providecommand \bibinfo  [0]{\@secondoftwo}%
\providecommand \bibfield  [0]{\@secondoftwo}%
\providecommand \translation [1]{[#1]}%
\providecommand \BibitemOpen [0]{}%
\providecommand \bibitemStop [0]{}%
\providecommand \bibitemNoStop [0]{.\EOS\space}%
\providecommand \EOS [0]{\spacefactor3000\relax}%
\providecommand \BibitemShut  [1]{\csname bibitem#1\endcsname}%
\let\auto@bib@innerbib\@empty
%</preamble>
\bibitem [{\citenamefont {Li}\ \emph {et~al.}(2014)\citenamefont {Li},
  \citenamefont {{van 't Erve}}, \citenamefont {Robinson}, \citenamefont {Liu},
  \citenamefont {Li},\ and\ \citenamefont {Jonker}}]{Li2014NN}%
  \BibitemOpen
  \bibfield  {author} {\bibinfo {author} {\bibfnamefont {C.~H.}\ \bibnamefont
  {Li}}, \bibinfo {author} {\bibfnamefont {O.~M.~J.}\ \bibnamefont {{van 't
  Erve}}}, \bibinfo {author} {\bibfnamefont {J.~T.}\ \bibnamefont {Robinson}},
  \bibinfo {author} {\bibfnamefont {Y.}~\bibnamefont {Liu}}, \bibinfo {author}
  {\bibfnamefont {L.}~\bibnamefont {Li}}, \ and\ \bibinfo {author}
  {\bibfnamefont {B.~T.}\ \bibnamefont {Jonker}},\ }\href {\doibase 10/gfz7vf}
  {\bibfield  {journal} {\bibinfo  {journal} {Nat.\ Nanotechnol.}\ }\textbf
  {\bibinfo {volume} {9}},\ \bibinfo {pages} {218} (\bibinfo {year}
  {2014})}\BibitemShut {NoStop}%
\bibitem [{\citenamefont {Ando}\ \emph {et~al.}(2014)\citenamefont {Ando},
  \citenamefont {Hamasaki}, \citenamefont {Kurokawa}, \citenamefont {Ichiba},
  \citenamefont {Yang}, \citenamefont {Novak}, \citenamefont {Sasaki},
  \citenamefont {Segawa}, \citenamefont {Ando},\ and\ \citenamefont
  {Shiraishi}}]{Ando2014NL}%
  \BibitemOpen
  \bibfield  {author} {\bibinfo {author} {\bibfnamefont {Y.}~\bibnamefont
  {Ando}}, \bibinfo {author} {\bibfnamefont {T.}~\bibnamefont {Hamasaki}},
  \bibinfo {author} {\bibfnamefont {T.}~\bibnamefont {Kurokawa}}, \bibinfo
  {author} {\bibfnamefont {K.}~\bibnamefont {Ichiba}}, \bibinfo {author}
  {\bibfnamefont {F.}~\bibnamefont {Yang}}, \bibinfo {author} {\bibfnamefont
  {M.}~\bibnamefont {Novak}}, \bibinfo {author} {\bibfnamefont
  {S.}~\bibnamefont {Sasaki}}, \bibinfo {author} {\bibfnamefont
  {K.}~\bibnamefont {Segawa}}, \bibinfo {author} {\bibfnamefont
  {Y.}~\bibnamefont {Ando}}, \ and\ \bibinfo {author} {\bibfnamefont
  {M.}~\bibnamefont {Shiraishi}},\ }\href {\doibase 10/f6rmpr} {\bibfield
  {journal} {\bibinfo  {journal} {Nano Lett.}\ }\textbf {\bibinfo {volume}
  {14}},\ \bibinfo {pages} {6226} (\bibinfo {year} {2014})}\BibitemShut
  {NoStop}%
\bibitem [{\citenamefont {DC}\ \emph {et~al.}(2018)\citenamefont {DC},
  \citenamefont {Grassi}, \citenamefont {Chen}, \citenamefont {Jamali},
  \citenamefont {Hickey}, \citenamefont {Zhang}, \citenamefont {Zhao},
  \citenamefont {Li}, \citenamefont {Quarterman}, \citenamefont {Lv},
  \citenamefont {Li}, \citenamefont {Manchon}, \citenamefont {Mkhoyan},
  \citenamefont {Low},\ and\ \citenamefont {Wang}}]{DC2018NM}%
  \BibitemOpen
  \bibfield  {author} {\bibinfo {author} {\bibfnamefont {M.}~\bibnamefont
  {DC}}, \bibinfo {author} {\bibfnamefont {R.}~\bibnamefont {Grassi}}, \bibinfo
  {author} {\bibfnamefont {J.-Y.}\ \bibnamefont {Chen}}, \bibinfo {author}
  {\bibfnamefont {M.}~\bibnamefont {Jamali}}, \bibinfo {author} {\bibfnamefont
  {D.~R.}\ \bibnamefont {Hickey}}, \bibinfo {author} {\bibfnamefont
  {D.}~\bibnamefont {Zhang}}, \bibinfo {author} {\bibfnamefont
  {Z.}~\bibnamefont {Zhao}}, \bibinfo {author} {\bibfnamefont {H.}~\bibnamefont
  {Li}}, \bibinfo {author} {\bibfnamefont {P.}~\bibnamefont {Quarterman}},
  \bibinfo {author} {\bibfnamefont {Y.}~\bibnamefont {Lv}}, \bibinfo {author}
  {\bibfnamefont {M.}~\bibnamefont {Li}}, \bibinfo {author} {\bibfnamefont
  {A.}~\bibnamefont {Manchon}}, \bibinfo {author} {\bibfnamefont {K.~A.}\
  \bibnamefont {Mkhoyan}}, \bibinfo {author} {\bibfnamefont {T.}~\bibnamefont
  {Low}}, \ and\ \bibinfo {author} {\bibfnamefont {J.-P.}\ \bibnamefont
  {Wang}},\ }\href {\doibase 10/gd69zx} {\bibfield  {journal} {\bibinfo
  {journal} {Nat.\ Mater.}\ }\textbf {\bibinfo {volume} {17}},\ \bibinfo
  {pages} {800} (\bibinfo {year} {2018})}\BibitemShut {NoStop}%
\bibitem [{\citenamefont {Kitaev}(2003)}]{Kitaev2003APb}%
  \BibitemOpen
  \bibfield  {author} {\bibinfo {author} {\bibfnamefont {A.~Y.}\ \bibnamefont
  {Kitaev}},\ }\href {\doibase 10/fvp239} {\bibfield  {journal} {\bibinfo
  {journal} {Ann.\ Phys.}\ }\textbf {\bibinfo {volume} {303}},\ \bibinfo
  {pages} {2} (\bibinfo {year} {2003})}\BibitemShut {NoStop}%
\bibitem [{\citenamefont {Kitagawa}\ \emph
  {et~al.}(2010{\natexlab{a}})\citenamefont {Kitagawa}, \citenamefont {Rudner},
  \citenamefont {Berg},\ and\ \citenamefont {Demler}}]{Kitagawa2010PRA}%
  \BibitemOpen
  \bibfield  {author} {\bibinfo {author} {\bibfnamefont {T.}~\bibnamefont
  {Kitagawa}}, \bibinfo {author} {\bibfnamefont {M.~S.}\ \bibnamefont
  {Rudner}}, \bibinfo {author} {\bibfnamefont {E.}~\bibnamefont {Berg}}, \ and\
  \bibinfo {author} {\bibfnamefont {E.}~\bibnamefont {Demler}},\ }\href
  {\doibase 10/fvfr7w} {\bibfield  {journal} {\bibinfo  {journal} {Phys.\ Rev.\
  A}\ }\textbf {\bibinfo {volume} {82}},\ \bibinfo {pages} {033429} (\bibinfo
  {year} {2010}{\natexlab{a}})}\BibitemShut {NoStop}%
\bibitem [{\citenamefont {Kitagawa}\ \emph
  {et~al.}(2010{\natexlab{b}})\citenamefont {Kitagawa}, \citenamefont {Berg},
  \citenamefont {Rudner},\ and\ \citenamefont {Demler}}]{Kitagawa2010PRB}%
  \BibitemOpen
  \bibfield  {author} {\bibinfo {author} {\bibfnamefont {T.}~\bibnamefont
  {Kitagawa}}, \bibinfo {author} {\bibfnamefont {E.}~\bibnamefont {Berg}},
  \bibinfo {author} {\bibfnamefont {M.}~\bibnamefont {Rudner}}, \ and\ \bibinfo
  {author} {\bibfnamefont {E.}~\bibnamefont {Demler}},\ }\href {\doibase
  10/b4hcz7} {\bibfield  {journal} {\bibinfo  {journal} {Phys.\ Rev.\ B}\
  }\textbf {\bibinfo {volume} {82}},\ \bibinfo {pages} {235114} (\bibinfo
  {year} {2010}{\natexlab{b}})}\BibitemShut {NoStop}%
\bibitem [{\citenamefont {Kitagawa}\ \emph {et~al.}(2012)\citenamefont
  {Kitagawa}, \citenamefont {Broome}, \citenamefont {Fedrizzi}, \citenamefont
  {Rudner}, \citenamefont {Berg}, \citenamefont {Kassal}, \citenamefont
  {{Aspuru-Guzik}}, \citenamefont {Demler},\ and\ \citenamefont
  {White}}]{Kitagawa2012NC}%
  \BibitemOpen
  \bibfield  {author} {\bibinfo {author} {\bibfnamefont {T.}~\bibnamefont
  {Kitagawa}}, \bibinfo {author} {\bibfnamefont {M.~A.}\ \bibnamefont
  {Broome}}, \bibinfo {author} {\bibfnamefont {A.}~\bibnamefont {Fedrizzi}},
  \bibinfo {author} {\bibfnamefont {M.~S.}\ \bibnamefont {Rudner}}, \bibinfo
  {author} {\bibfnamefont {E.}~\bibnamefont {Berg}}, \bibinfo {author}
  {\bibfnamefont {I.}~\bibnamefont {Kassal}}, \bibinfo {author} {\bibfnamefont
  {A.}~\bibnamefont {{Aspuru-Guzik}}}, \bibinfo {author} {\bibfnamefont
  {E.}~\bibnamefont {Demler}}, \ and\ \bibinfo {author} {\bibfnamefont {A.~G.}\
  \bibnamefont {White}},\ }\href {\doibase 10/f99837} {\bibfield  {journal}
  {\bibinfo  {journal} {Nat.\ Commun.}\ }\textbf {\bibinfo {volume} {3}},\
  \bibinfo {pages} {882} (\bibinfo {year} {2012})}\BibitemShut {NoStop}%
\bibitem [{\citenamefont {Asb\'oth}(2012)}]{Asboth2012PRB}%
  \BibitemOpen
  \bibfield  {author} {\bibinfo {author} {\bibfnamefont {J.~K.}\ \bibnamefont
  {Asb\'oth}},\ }\href {\doibase 10/gfxpb6} {\bibfield  {journal} {\bibinfo
  {journal} {Phys.\ Rev.\ B}\ }\textbf {\bibinfo {volume} {86}},\ \bibinfo
  {pages} {195414} (\bibinfo {year} {2012})}\BibitemShut {NoStop}%
\bibitem [{\citenamefont {Rudner}\ \emph {et~al.}(2013)\citenamefont {Rudner},
  \citenamefont {Lindner}, \citenamefont {Berg},\ and\ \citenamefont
  {Levin}}]{Rudner2013PRX}%
  \BibitemOpen
  \bibfield  {author} {\bibinfo {author} {\bibfnamefont {M.~S.}\ \bibnamefont
  {Rudner}}, \bibinfo {author} {\bibfnamefont {N.~H.}\ \bibnamefont {Lindner}},
  \bibinfo {author} {\bibfnamefont {E.}~\bibnamefont {Berg}}, \ and\ \bibinfo
  {author} {\bibfnamefont {M.}~\bibnamefont {Levin}},\ }\href {\doibase
  10/gfsjvb} {\bibfield  {journal} {\bibinfo  {journal} {Phys.\ Rev.\ X}\
  }\textbf {\bibinfo {volume} {3}},\ \bibinfo {pages} {031005} (\bibinfo {year}
  {2013})}\BibitemShut {NoStop}%
\bibitem [{\citenamefont {Zhang}\ \emph
  {et~al.}(2017{\natexlab{a}})\citenamefont {Zhang}, \citenamefont {Sanders},
  \citenamefont {Apers}, \citenamefont {Goyal},\ and\ \citenamefont
  {Feder}}]{Zhang2017PRL}%
  \BibitemOpen
  \bibfield  {author} {\bibinfo {author} {\bibfnamefont {W.-W.}\ \bibnamefont
  {Zhang}}, \bibinfo {author} {\bibfnamefont {B.~C.}\ \bibnamefont {Sanders}},
  \bibinfo {author} {\bibfnamefont {S.}~\bibnamefont {Apers}}, \bibinfo
  {author} {\bibfnamefont {S.~K.}\ \bibnamefont {Goyal}}, \ and\ \bibinfo
  {author} {\bibfnamefont {D.~L.}\ \bibnamefont {Feder}},\ }\href {\doibase
  10/gcpc5k} {\bibfield  {journal} {\bibinfo  {journal} {Phys.\ Rev.\ Lett.}\
  }\textbf {\bibinfo {volume} {119}},\ \bibinfo {pages} {197401} (\bibinfo
  {year} {2017}{\natexlab{a}})}\BibitemShut {NoStop}%
\bibitem [{\citenamefont {Xiao}\ \emph {et~al.}(2017)\citenamefont {Xiao},
  \citenamefont {Zhan}, \citenamefont {Bian}, \citenamefont {Wang},
  \citenamefont {Zhang}, \citenamefont {Wang}, \citenamefont {Li},
  \citenamefont {Mochizuki}, \citenamefont {Kim}, \citenamefont {Kawakami},
  \citenamefont {Yi}, \citenamefont {Obuse}, \citenamefont {Sanders},\ and\
  \citenamefont {Xue}}]{Xiao2017NP}%
  \BibitemOpen
  \bibfield  {author} {\bibinfo {author} {\bibfnamefont {L.}~\bibnamefont
  {Xiao}}, \bibinfo {author} {\bibfnamefont {X.}~\bibnamefont {Zhan}}, \bibinfo
  {author} {\bibfnamefont {Z.~H.}\ \bibnamefont {Bian}}, \bibinfo {author}
  {\bibfnamefont {K.~K.}\ \bibnamefont {Wang}}, \bibinfo {author}
  {\bibfnamefont {X.}~\bibnamefont {Zhang}}, \bibinfo {author} {\bibfnamefont
  {X.~P.}\ \bibnamefont {Wang}}, \bibinfo {author} {\bibfnamefont
  {J.}~\bibnamefont {Li}}, \bibinfo {author} {\bibfnamefont {K.}~\bibnamefont
  {Mochizuki}}, \bibinfo {author} {\bibfnamefont {D.}~\bibnamefont {Kim}},
  \bibinfo {author} {\bibfnamefont {N.}~\bibnamefont {Kawakami}}, \bibinfo
  {author} {\bibfnamefont {W.}~\bibnamefont {Yi}}, \bibinfo {author}
  {\bibfnamefont {H.}~\bibnamefont {Obuse}}, \bibinfo {author} {\bibfnamefont
  {B.~C.}\ \bibnamefont {Sanders}}, \ and\ \bibinfo {author} {\bibfnamefont
  {P.}~\bibnamefont {Xue}},\ }\href {\doibase 10/gcjdmt} {\bibfield  {journal}
  {\bibinfo  {journal} {Nat.\ Phys.}\ }\textbf {\bibinfo {volume} {13}},\
  \bibinfo {pages} {1117} (\bibinfo {year} {2017})}\BibitemShut {NoStop}%
\bibitem [{\citenamefont {Chen}\ \emph {et~al.}(2018)\citenamefont {Chen},
  \citenamefont {Ding}, \citenamefont {Qin}, \citenamefont {He}, \citenamefont
  {Luo}, \citenamefont {Chen}, \citenamefont {Liu}, \citenamefont {Wang},
  \citenamefont {Zhang}, \citenamefont {Li}, \citenamefont {You}, \citenamefont
  {Wang}, \citenamefont {Wang}, \citenamefont {Sanders}, \citenamefont {Lu},\
  and\ \citenamefont {Pan}}]{Chen2018PRL}%
  \BibitemOpen
  \bibfield  {author} {\bibinfo {author} {\bibfnamefont {C.}~\bibnamefont
  {Chen}}, \bibinfo {author} {\bibfnamefont {X.}~\bibnamefont {Ding}}, \bibinfo
  {author} {\bibfnamefont {J.}~\bibnamefont {Qin}}, \bibinfo {author}
  {\bibfnamefont {Y.}~\bibnamefont {He}}, \bibinfo {author} {\bibfnamefont
  {Y.-H.}\ \bibnamefont {Luo}}, \bibinfo {author} {\bibfnamefont {M.-C.}\
  \bibnamefont {Chen}}, \bibinfo {author} {\bibfnamefont {C.}~\bibnamefont
  {Liu}}, \bibinfo {author} {\bibfnamefont {X.-L.}\ \bibnamefont {Wang}},
  \bibinfo {author} {\bibfnamefont {W.-J.}\ \bibnamefont {Zhang}}, \bibinfo
  {author} {\bibfnamefont {H.}~\bibnamefont {Li}}, \bibinfo {author}
  {\bibfnamefont {L.-X.}\ \bibnamefont {You}}, \bibinfo {author} {\bibfnamefont
  {Z.}~\bibnamefont {Wang}}, \bibinfo {author} {\bibfnamefont {D.-W.}\
  \bibnamefont {Wang}}, \bibinfo {author} {\bibfnamefont {B.~C.}\ \bibnamefont
  {Sanders}}, \bibinfo {author} {\bibfnamefont {C.-Y.}\ \bibnamefont {Lu}}, \
  and\ \bibinfo {author} {\bibfnamefont {J.-W.}\ \bibnamefont {Pan}},\ }\href
  {\doibase 10/gd55m3} {\bibfield  {journal} {\bibinfo  {journal} {Phys.\ Rev.\
  Lett.}\ }\textbf {\bibinfo {volume} {121}},\ \bibinfo {pages} {100502}
  (\bibinfo {year} {2018})}\BibitemShut {NoStop}%
\bibitem [{\citenamefont {Sun}\ \emph {et~al.}(2018)\citenamefont {Sun},
  \citenamefont {Yi}, \citenamefont {Wang}, \citenamefont {Zhang},
  \citenamefont {Sanders}, \citenamefont {Xu}, \citenamefont {Wang},
  \citenamefont {Schmiedmayer}, \citenamefont {Deng}, \citenamefont {Liu},
  \citenamefont {Chen},\ and\ \citenamefont {Pan}}]{Sun2018PRL}%
  \BibitemOpen
  \bibfield  {author} {\bibinfo {author} {\bibfnamefont {W.}~\bibnamefont
  {Sun}}, \bibinfo {author} {\bibfnamefont {C.-R.}\ \bibnamefont {Yi}},
  \bibinfo {author} {\bibfnamefont {B.-Z.}\ \bibnamefont {Wang}}, \bibinfo
  {author} {\bibfnamefont {W.-W.}\ \bibnamefont {Zhang}}, \bibinfo {author}
  {\bibfnamefont {B.~C.}\ \bibnamefont {Sanders}}, \bibinfo {author}
  {\bibfnamefont {X.-T.}\ \bibnamefont {Xu}}, \bibinfo {author} {\bibfnamefont
  {Z.-Y.}\ \bibnamefont {Wang}}, \bibinfo {author} {\bibfnamefont
  {J.}~\bibnamefont {Schmiedmayer}}, \bibinfo {author} {\bibfnamefont
  {Y.}~\bibnamefont {Deng}}, \bibinfo {author} {\bibfnamefont {X.-J.}\
  \bibnamefont {Liu}}, \bibinfo {author} {\bibfnamefont {S.}~\bibnamefont
  {Chen}}, \ and\ \bibinfo {author} {\bibfnamefont {J.-W.}\ \bibnamefont
  {Pan}},\ }\href {\doibase 10/gfr93s} {\bibfield  {journal} {\bibinfo
  {journal} {Phys.\ Rev.\ Lett.}\ }\textbf {\bibinfo {volume} {121}},\ \bibinfo
  {pages} {250403} (\bibinfo {year} {2018})}\BibitemShut {NoStop}%
\bibitem [{\citenamefont {Sajid}\ \emph {et~al.}(2019)\citenamefont {Sajid},
  \citenamefont {Asb\'oth}, \citenamefont {Meschede}, \citenamefont {Werner},\
  and\ \citenamefont {Alberti}}]{Sajid2019PRB}%
  \BibitemOpen
  \bibfield  {author} {\bibinfo {author} {\bibfnamefont {M.}~\bibnamefont
  {Sajid}}, \bibinfo {author} {\bibfnamefont {J.~K.}\ \bibnamefont {Asb\'oth}},
  \bibinfo {author} {\bibfnamefont {D.}~\bibnamefont {Meschede}}, \bibinfo
  {author} {\bibfnamefont {R.~F.}\ \bibnamefont {Werner}}, \ and\ \bibinfo
  {author} {\bibfnamefont {A.}~\bibnamefont {Alberti}},\ }\href {\doibase
  10.1103/PhysRevB.99.214303} {\bibfield  {journal} {\bibinfo  {journal}
  {Phys.\ Rev.\ B}\ }\textbf {\bibinfo {volume} {99}},\ \bibinfo {pages}
  {214303} (\bibinfo {year} {2019})}\BibitemShut {NoStop}%
\bibitem [{\citenamefont {Kempe}(2003)}]{Kempe2003CP}%
  \BibitemOpen
  \bibfield  {author} {\bibinfo {author} {\bibfnamefont {J.}~\bibnamefont
  {Kempe}},\ }\href {\doibase 10/bpftc8} {\bibfield  {journal} {\bibinfo
  {journal} {Contemp. Phys.}\ }\textbf {\bibinfo {volume} {44}},\ \bibinfo
  {pages} {307} (\bibinfo {year} {2003})}\BibitemShut {NoStop}%
\bibitem [{\citenamefont {Aharonov}\ \emph {et~al.}(1993)\citenamefont
  {Aharonov}, \citenamefont {Davidovich},\ and\ \citenamefont
  {Zagury}}]{Aharonov1993PRA}%
  \BibitemOpen
  \bibfield  {author} {\bibinfo {author} {\bibfnamefont {Y.}~\bibnamefont
  {Aharonov}}, \bibinfo {author} {\bibfnamefont {L.}~\bibnamefont
  {Davidovich}}, \ and\ \bibinfo {author} {\bibfnamefont {N.}~\bibnamefont
  {Zagury}},\ }\href {\doibase 10/dg9qdz} {\bibfield  {journal} {\bibinfo
  {journal} {Phys.\ Rev.\ A}\ }\textbf {\bibinfo {volume} {48}},\ \bibinfo
  {pages} {1687} (\bibinfo {year} {1993})}\BibitemShut {NoStop}%
\bibitem [{\citenamefont {Aharonov}\ \emph {et~al.}(2001)\citenamefont
  {Aharonov}, \citenamefont {Ambainis}, \citenamefont {Kempe},\ and\
  \citenamefont {Vazirani}}]{Aharonov2001STOC}%
  \BibitemOpen
  \bibfield  {author} {\bibinfo {author} {\bibfnamefont {D.}~\bibnamefont
  {Aharonov}}, \bibinfo {author} {\bibfnamefont {A.}~\bibnamefont {Ambainis}},
  \bibinfo {author} {\bibfnamefont {J.}~\bibnamefont {Kempe}}, \ and\ \bibinfo
  {author} {\bibfnamefont {U.}~\bibnamefont {Vazirani}},\ }in\ \href {\doibase
  10/fppcc6} {\emph {\bibinfo {booktitle} {Proc.\ 33rd Annual ACM Symposium on
  Theory of Computing}}}\ (\bibinfo  {publisher} {ACM},\ \bibinfo {address}
  {New York},\ \bibinfo {year} {2001})\ pp.\ \bibinfo {pages}
  {50--59}\BibitemShut {NoStop}%
\bibitem [{\citenamefont {Ambainis}\ \emph {et~al.}(2001)\citenamefont
  {Ambainis}, \citenamefont {Bach}, \citenamefont {Nayak}, \citenamefont
  {Vishwanath},\ and\ \citenamefont {Watrous}}]{Ambainis2001STOC}%
  \BibitemOpen
  \bibfield  {author} {\bibinfo {author} {\bibfnamefont {A.}~\bibnamefont
  {Ambainis}}, \bibinfo {author} {\bibfnamefont {E.}~\bibnamefont {Bach}},
  \bibinfo {author} {\bibfnamefont {A.}~\bibnamefont {Nayak}}, \bibinfo
  {author} {\bibfnamefont {A.}~\bibnamefont {Vishwanath}}, \ and\ \bibinfo
  {author} {\bibfnamefont {J.}~\bibnamefont {Watrous}},\ }in\ \href {\doibase
  10.1145/380752.380757} {\emph {\bibinfo {booktitle} {Proc.\ 33rd Annual ACM
  Symposium on Theory of Computing}}}\ (\bibinfo  {publisher} {ACM},\ \bibinfo
  {address} {New York},\ \bibinfo {year} {2001})\ pp.\ \bibinfo {pages}
  {37--49}\BibitemShut {NoStop}%
\bibitem [{\citenamefont {Mackay}\ \emph {et~al.}(2002)\citenamefont {Mackay},
  \citenamefont {Bartlett}, \citenamefont {Stephenson},\ and\ \citenamefont
  {Sanders}}]{Mackay2002JPA}%
  \BibitemOpen
  \bibfield  {author} {\bibinfo {author} {\bibfnamefont {T.~D.}\ \bibnamefont
  {Mackay}}, \bibinfo {author} {\bibfnamefont {S.~D.}\ \bibnamefont
  {Bartlett}}, \bibinfo {author} {\bibfnamefont {L.~T.}\ \bibnamefont
  {Stephenson}}, \ and\ \bibinfo {author} {\bibfnamefont {B.~C.}\ \bibnamefont
  {Sanders}},\ }\href {\doibase 10/fhqjx2} {\bibfield  {journal} {\bibinfo
  {journal} {J.\ Phys.\ A}\ }\textbf {\bibinfo {volume} {35}},\ \bibinfo
  {pages} {2745} (\bibinfo {year} {2002})}\BibitemShut {NoStop}%
\bibitem [{\citenamefont {Childs}\ \emph {et~al.}(2003)\citenamefont {Childs},
  \citenamefont {Cleve}, \citenamefont {Deotto}, \citenamefont {Farhi},
  \citenamefont {Gutmann},\ and\ \citenamefont {Spielman}}]{Childs2003STOC}%
  \BibitemOpen
  \bibfield  {author} {\bibinfo {author} {\bibfnamefont {A.~M.}\ \bibnamefont
  {Childs}}, \bibinfo {author} {\bibfnamefont {R.}~\bibnamefont {Cleve}},
  \bibinfo {author} {\bibfnamefont {E.}~\bibnamefont {Deotto}}, \bibinfo
  {author} {\bibfnamefont {E.}~\bibnamefont {Farhi}}, \bibinfo {author}
  {\bibfnamefont {S.}~\bibnamefont {Gutmann}}, \ and\ \bibinfo {author}
  {\bibfnamefont {D.~A.}\ \bibnamefont {Spielman}},\ }in\ \href {\doibase
  10/fdr82p} {\emph {\bibinfo {booktitle} {Proc.\ 35th Annual ACM Symposium on
  Theory of Computing}}}\ (\bibinfo  {publisher} {ACM},\ \bibinfo {address}
  {New York},\ \bibinfo {year} {2003})\ pp.\ \bibinfo {pages}
  {59--68}\BibitemShut {NoStop}%
\bibitem [{\citenamefont {Ambainis}(2007)}]{Ambainis2007SJC}%
  \BibitemOpen
  \bibfield  {author} {\bibinfo {author} {\bibfnamefont {A.}~\bibnamefont
  {Ambainis}},\ }\href {\doibase 10/b2mvvf} {\bibfield  {journal} {\bibinfo
  {journal} {SIAM J.\ Comput.}\ }\textbf {\bibinfo {volume} {37}},\ \bibinfo
  {pages} {210} (\bibinfo {year} {2007})}\BibitemShut {NoStop}%
\bibitem [{\citenamefont {Magniez}\ \emph {et~al.}(2007)\citenamefont
  {Magniez}, \citenamefont {Santha},\ and\ \citenamefont
  {Szegedy}}]{Magniez2007SJC}%
  \BibitemOpen
  \bibfield  {author} {\bibinfo {author} {\bibfnamefont {F.}~\bibnamefont
  {Magniez}}, \bibinfo {author} {\bibfnamefont {M.}~\bibnamefont {Santha}}, \
  and\ \bibinfo {author} {\bibfnamefont {M.}~\bibnamefont {Szegedy}},\ }\href
  {\doibase 10/fmj262} {\bibfield  {journal} {\bibinfo  {journal} {SIAM J.\
  Comput.}\ }\textbf {\bibinfo {volume} {37}},\ \bibinfo {pages} {413}
  (\bibinfo {year} {2007})}\BibitemShut {NoStop}%
\bibitem [{\citenamefont {Farhi}\ \emph {et~al.}(2008)\citenamefont {Farhi},
  \citenamefont {Goldstone},\ and\ \citenamefont {Gutmann}}]{Farhi2008TC}%
  \BibitemOpen
  \bibfield  {author} {\bibinfo {author} {\bibfnamefont {E.}~\bibnamefont
  {Farhi}}, \bibinfo {author} {\bibfnamefont {J.}~\bibnamefont {Goldstone}}, \
  and\ \bibinfo {author} {\bibfnamefont {S.}~\bibnamefont {Gutmann}},\ }\href
  {\doibase 10/fckcn2} {\bibfield  {journal} {\bibinfo  {journal} {Theory
  Comput.}\ }\textbf {\bibinfo {volume} {4}},\ \bibinfo {pages} {169} (\bibinfo
  {year} {2008})}\BibitemShut {NoStop}%
\bibitem [{\citenamefont {Douglas}\ and\ \citenamefont
  {Wang}(2008)}]{Douglas2008JPA}%
  \BibitemOpen
  \bibfield  {author} {\bibinfo {author} {\bibfnamefont {B.~L.}\ \bibnamefont
  {Douglas}}\ and\ \bibinfo {author} {\bibfnamefont {J.}~\bibnamefont {Wang}},\
  }\href {\doibase 10/cdgzrz} {\bibfield  {journal} {\bibinfo  {journal} {J.
  Phys. A}\ }\textbf {\bibinfo {volume} {41}},\ \bibinfo {pages} {075303}
  (\bibinfo {year} {2008})}\BibitemShut {NoStop}%
\bibitem [{\citenamefont {Childs}(2009)}]{Childs2009PRL}%
  \BibitemOpen
  \bibfield  {author} {\bibinfo {author} {\bibfnamefont {A.~M.}\ \bibnamefont
  {Childs}},\ }\href {\doibase 10/fb5k7b} {\bibfield  {journal} {\bibinfo
  {journal} {Phys.\ Rev.\ Lett.}\ }\textbf {\bibinfo {volume} {102}},\ \bibinfo
  {pages} {180501} (\bibinfo {year} {2009})}\BibitemShut {NoStop}%
\bibitem [{\citenamefont {Godoy}\ and\ \citenamefont
  {Fujita}(1992)}]{Godoy1992JCP}%
  \BibitemOpen
  \bibfield  {author} {\bibinfo {author} {\bibfnamefont {S.}~\bibnamefont
  {Godoy}}\ and\ \bibinfo {author} {\bibfnamefont {S.}~\bibnamefont {Fujita}},\
  }\href {\doibase 10.1063/1.463812} {\bibfield  {journal} {\bibinfo  {journal}
  {J.\ Chem.\ Phys.}\ }\textbf {\bibinfo {volume} {97}},\ \bibinfo {pages}
  {5148} (\bibinfo {year} {1992})}\BibitemShut {NoStop}%
\bibitem [{\citenamefont {M\"ulken}\ and\ \citenamefont
  {Blumen}(2011)}]{Mulken2011PR}%
  \BibitemOpen
  \bibfield  {author} {\bibinfo {author} {\bibfnamefont {O.}~\bibnamefont
  {M\"ulken}}\ and\ \bibinfo {author} {\bibfnamefont {A.}~\bibnamefont
  {Blumen}},\ }\href {\doibase 10.1016/j.physrep.2011.01.002} {\bibfield
  {journal} {\bibinfo  {journal} {Phys.\ Rep.}\ }\textbf {\bibinfo {volume}
  {502}},\ \bibinfo {pages} {37} (\bibinfo {year} {2011})}\BibitemShut
  {NoStop}%
\bibitem [{\citenamefont {Di~Molfetta}\ \emph {et~al.}(2013)\citenamefont
  {Di~Molfetta}, \citenamefont {Brachet},\ and\ \citenamefont
  {Debbasch}}]{DiMolfetta2013PRA}%
  \BibitemOpen
  \bibfield  {author} {\bibinfo {author} {\bibfnamefont {G.}~\bibnamefont
  {Di~Molfetta}}, \bibinfo {author} {\bibfnamefont {M.}~\bibnamefont
  {Brachet}}, \ and\ \bibinfo {author} {\bibfnamefont {F.}~\bibnamefont
  {Debbasch}},\ }\href {\doibase 10/gf5gwb} {\bibfield  {journal} {\bibinfo
  {journal} {Phys.\ Rev.\ A}\ }\textbf {\bibinfo {volume} {88}},\ \bibinfo
  {pages} {042301} (\bibinfo {year} {2013})}\BibitemShut {NoStop}%
\bibitem [{\citenamefont {Zhang}\ \emph {et~al.}(2016)\citenamefont {Zhang},
  \citenamefont {Goyal}, \citenamefont {Gao}, \citenamefont {Sanders},\ and\
  \citenamefont {Simon}}]{Zhang2016NJP}%
  \BibitemOpen
  \bibfield  {author} {\bibinfo {author} {\bibfnamefont {W.-W.}\ \bibnamefont
  {Zhang}}, \bibinfo {author} {\bibfnamefont {S.~K.}\ \bibnamefont {Goyal}},
  \bibinfo {author} {\bibfnamefont {F.}~\bibnamefont {Gao}}, \bibinfo {author}
  {\bibfnamefont {B.~C.}\ \bibnamefont {Sanders}}, \ and\ \bibinfo {author}
  {\bibfnamefont {C.}~\bibnamefont {Simon}},\ }\href {\doibase 10/gd7fb2}
  {\bibfield  {journal} {\bibinfo  {journal} {New J.\ Phys.}\ }\textbf
  {\bibinfo {volume} {18}},\ \bibinfo {pages} {093025} (\bibinfo {year}
  {2016})}\BibitemShut {NoStop}%
\bibitem [{\citenamefont {Schmitz}\ \emph {et~al.}(2009)\citenamefont
  {Schmitz}, \citenamefont {Matjeschk}, \citenamefont {Schneider},
  \citenamefont {Glueckert}, \citenamefont {Enderlein}, \citenamefont {Huber},\
  and\ \citenamefont {Schaetz}}]{Schmitz2009PRL}%
  \BibitemOpen
  \bibfield  {author} {\bibinfo {author} {\bibfnamefont {H.}~\bibnamefont
  {Schmitz}}, \bibinfo {author} {\bibfnamefont {R.}~\bibnamefont {Matjeschk}},
  \bibinfo {author} {\bibfnamefont {C.}~\bibnamefont {Schneider}}, \bibinfo
  {author} {\bibfnamefont {J.}~\bibnamefont {Glueckert}}, \bibinfo {author}
  {\bibfnamefont {M.}~\bibnamefont {Enderlein}}, \bibinfo {author}
  {\bibfnamefont {T.}~\bibnamefont {Huber}}, \ and\ \bibinfo {author}
  {\bibfnamefont {T.}~\bibnamefont {Schaetz}},\ }\href {\doibase
  10.1103/PhysRevLett.103.090504} {\bibfield  {journal} {\bibinfo  {journal}
  {Phys.\ Rev.\ Lett.}\ }\textbf {\bibinfo {volume} {103}},\ \bibinfo {pages}
  {090504} (\bibinfo {year} {2009})}\BibitemShut {NoStop}%
\bibitem [{\citenamefont {Z\"ahringer}\ \emph {et~al.}(2010)\citenamefont
  {Z\"ahringer}, \citenamefont {Kirchmair}, \citenamefont {Gerritsma},
  \citenamefont {Solano}, \citenamefont {Blatt},\ and\ \citenamefont
  {Roos}}]{Zahringer2010PRL}%
  \BibitemOpen
  \bibfield  {author} {\bibinfo {author} {\bibfnamefont {F.}~\bibnamefont
  {Z\"ahringer}}, \bibinfo {author} {\bibfnamefont {G.}~\bibnamefont
  {Kirchmair}}, \bibinfo {author} {\bibfnamefont {R.}~\bibnamefont
  {Gerritsma}}, \bibinfo {author} {\bibfnamefont {E.}~\bibnamefont {Solano}},
  \bibinfo {author} {\bibfnamefont {R.}~\bibnamefont {Blatt}}, \ and\ \bibinfo
  {author} {\bibfnamefont {C.~F.}\ \bibnamefont {Roos}},\ }\href {\doibase
  10.1103/PhysRevLett.104.100503} {\bibfield  {journal} {\bibinfo  {journal}
  {Phys.\ Rev.\ Lett.}\ }\textbf {\bibinfo {volume} {104}},\ \bibinfo {pages}
  {100503} (\bibinfo {year} {2010})}\BibitemShut {NoStop}%
\bibitem [{\citenamefont {Flurin}\ \emph {et~al.}(2017)\citenamefont {Flurin},
  \citenamefont {Ramasesh}, \citenamefont {{Hacohen-Gourgy}}, \citenamefont
  {Martin}, \citenamefont {Yao},\ and\ \citenamefont
  {Siddiqi}}]{Flurin2017PRX}%
  \BibitemOpen
  \bibfield  {author} {\bibinfo {author} {\bibfnamefont {E.}~\bibnamefont
  {Flurin}}, \bibinfo {author} {\bibfnamefont {V.~V.}\ \bibnamefont
  {Ramasesh}}, \bibinfo {author} {\bibfnamefont {S.}~\bibnamefont
  {{Hacohen-Gourgy}}}, \bibinfo {author} {\bibfnamefont {L.~S.}\ \bibnamefont
  {Martin}}, \bibinfo {author} {\bibfnamefont {N.~Y.}\ \bibnamefont {Yao}}, \
  and\ \bibinfo {author} {\bibfnamefont {I.}~\bibnamefont {Siddiqi}},\ }\href
  {\doibase 10.1103/PhysRevX.7.031023} {\bibfield  {journal} {\bibinfo
  {journal} {Phys.\ Rev.\ X}\ }\textbf {\bibinfo {volume} {7}},\ \bibinfo
  {pages} {031023} (\bibinfo {year} {2017})}\BibitemShut {NoStop}%
\bibitem [{\citenamefont {Yan}\ \emph {et~al.}(2019)\citenamefont {Yan},
  \citenamefont {Zhang}, \citenamefont {Gong}, \citenamefont {Wu},
  \citenamefont {Zheng}, \citenamefont {Li}, \citenamefont {Wang},
  \citenamefont {Liang}, \citenamefont {Lin}, \citenamefont {Xu}, \citenamefont
  {Guo}, \citenamefont {Sun}, \citenamefont {Peng}, \citenamefont {Xia},
  \citenamefont {Deng}, \citenamefont {Rong}, \citenamefont {You},
  \citenamefont {Nori}, \citenamefont {Fan}, \citenamefont {Zhu},\ and\
  \citenamefont {Pan}}]{Yan2019Science}%
  \BibitemOpen
  \bibfield  {author} {\bibinfo {author} {\bibfnamefont {Z.}~\bibnamefont
  {Yan}}, \bibinfo {author} {\bibfnamefont {Y.-R.}\ \bibnamefont {Zhang}},
  \bibinfo {author} {\bibfnamefont {M.}~\bibnamefont {Gong}}, \bibinfo {author}
  {\bibfnamefont {Y.}~\bibnamefont {Wu}}, \bibinfo {author} {\bibfnamefont
  {Y.}~\bibnamefont {Zheng}}, \bibinfo {author} {\bibfnamefont
  {S.}~\bibnamefont {Li}}, \bibinfo {author} {\bibfnamefont {C.}~\bibnamefont
  {Wang}}, \bibinfo {author} {\bibfnamefont {F.}~\bibnamefont {Liang}},
  \bibinfo {author} {\bibfnamefont {J.}~\bibnamefont {Lin}}, \bibinfo {author}
  {\bibfnamefont {Y.}~\bibnamefont {Xu}}, \bibinfo {author} {\bibfnamefont
  {C.}~\bibnamefont {Guo}}, \bibinfo {author} {\bibfnamefont {L.}~\bibnamefont
  {Sun}}, \bibinfo {author} {\bibfnamefont {C.-Z.}\ \bibnamefont {Peng}},
  \bibinfo {author} {\bibfnamefont {K.}~\bibnamefont {Xia}}, \bibinfo {author}
  {\bibfnamefont {H.}~\bibnamefont {Deng}}, \bibinfo {author} {\bibfnamefont
  {H.}~\bibnamefont {Rong}}, \bibinfo {author} {\bibfnamefont {J.~Q.}\
  \bibnamefont {You}}, \bibinfo {author} {\bibfnamefont {F.}~\bibnamefont
  {Nori}}, \bibinfo {author} {\bibfnamefont {H.}~\bibnamefont {Fan}}, \bibinfo
  {author} {\bibfnamefont {X.}~\bibnamefont {Zhu}}, \ and\ \bibinfo {author}
  {\bibfnamefont {J.-W.}\ \bibnamefont {Pan}},\ }\href {\doibase 10/gf2bnb}
  {\bibfield  {journal} {\bibinfo  {journal} {Science}\ }\textbf {\bibinfo
  {volume} {364}},\ \bibinfo {pages} {753} (\bibinfo {year}
  {2019})}\BibitemShut {NoStop}%
\bibitem [{\citenamefont {Du}\ \emph {et~al.}(2003)\citenamefont {Du},
  \citenamefont {Li}, \citenamefont {Xu}, \citenamefont {Shi}, \citenamefont
  {Wu}, \citenamefont {Zhou},\ and\ \citenamefont {Han}}]{Du2003PRA}%
  \BibitemOpen
  \bibfield  {author} {\bibinfo {author} {\bibfnamefont {J.}~\bibnamefont
  {Du}}, \bibinfo {author} {\bibfnamefont {H.}~\bibnamefont {Li}}, \bibinfo
  {author} {\bibfnamefont {X.}~\bibnamefont {Xu}}, \bibinfo {author}
  {\bibfnamefont {M.}~\bibnamefont {Shi}}, \bibinfo {author} {\bibfnamefont
  {J.}~\bibnamefont {Wu}}, \bibinfo {author} {\bibfnamefont {X.}~\bibnamefont
  {Zhou}}, \ and\ \bibinfo {author} {\bibfnamefont {R.}~\bibnamefont {Han}},\
  }\href {\doibase 10/dhzmzk} {\bibfield  {journal} {\bibinfo  {journal}
  {Phys.\ Rev.\ A}\ }\textbf {\bibinfo {volume} {67}},\ \bibinfo {pages}
  {042316} (\bibinfo {year} {2003})}\BibitemShut {NoStop}%
\bibitem [{\citenamefont {Ryan}\ \emph {et~al.}(2005)\citenamefont {Ryan},
  \citenamefont {Laforest}, \citenamefont {Boileau},\ and\ \citenamefont
  {Laflamme}}]{Ryan2005PRA}%
  \BibitemOpen
  \bibfield  {author} {\bibinfo {author} {\bibfnamefont {C.~A.}\ \bibnamefont
  {Ryan}}, \bibinfo {author} {\bibfnamefont {M.}~\bibnamefont {Laforest}},
  \bibinfo {author} {\bibfnamefont {J.~C.}\ \bibnamefont {Boileau}}, \ and\
  \bibinfo {author} {\bibfnamefont {R.}~\bibnamefont {Laflamme}},\ }\href
  {\doibase 10/fdx2bh} {\bibfield  {journal} {\bibinfo  {journal} {Phys.\ Rev.\
  A}\ }\textbf {\bibinfo {volume} {72}},\ \bibinfo {pages} {062317} (\bibinfo
  {year} {2005})}\BibitemShut {NoStop}%
\bibitem [{\citenamefont {Karski}\ \emph {et~al.}(2009)\citenamefont {Karski},
  \citenamefont {F\"orster}, \citenamefont {Choi}, \citenamefont {Steffen},
  \citenamefont {Alt}, \citenamefont {Meschede},\ and\ \citenamefont
  {Widera}}]{Karski2009Science}%
  \BibitemOpen
  \bibfield  {author} {\bibinfo {author} {\bibfnamefont {M.}~\bibnamefont
  {Karski}}, \bibinfo {author} {\bibfnamefont {L.}~\bibnamefont {F\"orster}},
  \bibinfo {author} {\bibfnamefont {J.-M.}\ \bibnamefont {Choi}}, \bibinfo
  {author} {\bibfnamefont {A.}~\bibnamefont {Steffen}}, \bibinfo {author}
  {\bibfnamefont {W.}~\bibnamefont {Alt}}, \bibinfo {author} {\bibfnamefont
  {D.}~\bibnamefont {Meschede}}, \ and\ \bibinfo {author} {\bibfnamefont
  {A.}~\bibnamefont {Widera}},\ }\href {\doibase 10.1126/science.1174436}
  {\bibfield  {journal} {\bibinfo  {journal} {Science}\ }\textbf {\bibinfo
  {volume} {325}},\ \bibinfo {pages} {174} (\bibinfo {year}
  {2009})}\BibitemShut {NoStop}%
\bibitem [{\citenamefont {Dadras}\ \emph {et~al.}(2018)\citenamefont {Dadras},
  \citenamefont {Gresch}, \citenamefont {Groiseau}, \citenamefont {Wimberger},\
  and\ \citenamefont {Summy}}]{Dadras2018PRL}%
  \BibitemOpen
  \bibfield  {author} {\bibinfo {author} {\bibfnamefont {S.}~\bibnamefont
  {Dadras}}, \bibinfo {author} {\bibfnamefont {A.}~\bibnamefont {Gresch}},
  \bibinfo {author} {\bibfnamefont {C.}~\bibnamefont {Groiseau}}, \bibinfo
  {author} {\bibfnamefont {S.}~\bibnamefont {Wimberger}}, \ and\ \bibinfo
  {author} {\bibfnamefont {G.~S.}\ \bibnamefont {Summy}},\ }\href {\doibase
  10/gd2xsx} {\bibfield  {journal} {\bibinfo  {journal} {Phys.\ Rev.\ Lett.}\
  }\textbf {\bibinfo {volume} {121}},\ \bibinfo {pages} {070402} (\bibinfo
  {year} {2018})}\BibitemShut {NoStop}%
\bibitem [{\citenamefont {Do}\ \emph {et~al.}(2005)\citenamefont {Do},
  \citenamefont {Stohler}, \citenamefont {Balasubramanian}, \citenamefont
  {Elliott}, \citenamefont {Eash}, \citenamefont {Fischbach}, \citenamefont
  {Fischbach}, \citenamefont {Mills},\ and\ \citenamefont
  {Zwickl}}]{Do2005JOSAB}%
  \BibitemOpen
  \bibfield  {author} {\bibinfo {author} {\bibfnamefont {B.}~\bibnamefont
  {Do}}, \bibinfo {author} {\bibfnamefont {M.~L.}\ \bibnamefont {Stohler}},
  \bibinfo {author} {\bibfnamefont {S.}~\bibnamefont {Balasubramanian}},
  \bibinfo {author} {\bibfnamefont {D.~S.}\ \bibnamefont {Elliott}}, \bibinfo
  {author} {\bibfnamefont {C.}~\bibnamefont {Eash}}, \bibinfo {author}
  {\bibfnamefont {E.}~\bibnamefont {Fischbach}}, \bibinfo {author}
  {\bibfnamefont {M.~A.}\ \bibnamefont {Fischbach}}, \bibinfo {author}
  {\bibfnamefont {A.}~\bibnamefont {Mills}}, \ and\ \bibinfo {author}
  {\bibfnamefont {B.}~\bibnamefont {Zwickl}},\ }\href {\doibase
  10.1364/JOSAB.22.000499} {\bibfield  {journal} {\bibinfo  {journal} {J.\
  Opt.\ Soc.\ Am.\ B}\ }\textbf {\bibinfo {volume} {22}},\ \bibinfo {pages}
  {499} (\bibinfo {year} {2005})}\BibitemShut {NoStop}%
\bibitem [{\citenamefont {Cardano}\ \emph {et~al.}(2015)\citenamefont
  {Cardano}, \citenamefont {Massa}, \citenamefont {Qassim}, \citenamefont
  {Karimi}, \citenamefont {Slussarenko}, \citenamefont {Paparo}, \citenamefont
  {de~Lisio}, \citenamefont {Sciarrino}, \citenamefont {Santamato},
  \citenamefont {Boyd},\ and\ \citenamefont {Marrucci}}]{Cardano2015SA}%
  \BibitemOpen
  \bibfield  {author} {\bibinfo {author} {\bibfnamefont {F.}~\bibnamefont
  {Cardano}}, \bibinfo {author} {\bibfnamefont {F.}~\bibnamefont {Massa}},
  \bibinfo {author} {\bibfnamefont {H.}~\bibnamefont {Qassim}}, \bibinfo
  {author} {\bibfnamefont {E.}~\bibnamefont {Karimi}}, \bibinfo {author}
  {\bibfnamefont {S.}~\bibnamefont {Slussarenko}}, \bibinfo {author}
  {\bibfnamefont {D.}~\bibnamefont {Paparo}}, \bibinfo {author} {\bibfnamefont
  {C.}~\bibnamefont {de~Lisio}}, \bibinfo {author} {\bibfnamefont
  {F.}~\bibnamefont {Sciarrino}}, \bibinfo {author} {\bibfnamefont
  {E.}~\bibnamefont {Santamato}}, \bibinfo {author} {\bibfnamefont {R.~W.}\
  \bibnamefont {Boyd}}, \ and\ \bibinfo {author} {\bibfnamefont
  {L.}~\bibnamefont {Marrucci}},\ }\href {\doibase 10.1126/sciadv.1500087}
  {\bibfield  {journal} {\bibinfo  {journal} {Sci.\ Adv.}\ }\textbf {\bibinfo
  {volume} {1}},\ \bibinfo {pages} {e1500087} (\bibinfo {year}
  {2015})}\BibitemShut {NoStop}%
\bibitem [{\citenamefont {Perets}\ \emph {et~al.}(2008)\citenamefont {Perets},
  \citenamefont {Lahini}, \citenamefont {Pozzi}, \citenamefont {Sorel},
  \citenamefont {Morandotti},\ and\ \citenamefont
  {Silberberg}}]{Perets2008PRL}%
  \BibitemOpen
  \bibfield  {author} {\bibinfo {author} {\bibfnamefont {H.~B.}\ \bibnamefont
  {Perets}}, \bibinfo {author} {\bibfnamefont {Y.}~\bibnamefont {Lahini}},
  \bibinfo {author} {\bibfnamefont {F.}~\bibnamefont {Pozzi}}, \bibinfo
  {author} {\bibfnamefont {M.}~\bibnamefont {Sorel}}, \bibinfo {author}
  {\bibfnamefont {R.}~\bibnamefont {Morandotti}}, \ and\ \bibinfo {author}
  {\bibfnamefont {Y.}~\bibnamefont {Silberberg}},\ }\href {\doibase 10/cxtnjt}
  {\bibfield  {journal} {\bibinfo  {journal} {Phys. Rev. Lett.}\ }\textbf
  {\bibinfo {volume} {100}},\ \bibinfo {pages} {170506} (\bibinfo {year}
  {2008})}\BibitemShut {NoStop}%
\bibitem [{\citenamefont {Tang}\ \emph {et~al.}(2018)\citenamefont {Tang},
  \citenamefont {Lin}, \citenamefont {Feng}, \citenamefont {Chen},
  \citenamefont {Gao}, \citenamefont {Sun}, \citenamefont {Wang}, \citenamefont
  {Lai}, \citenamefont {Xu}, \citenamefont {Wang}, \citenamefont {Qiao},
  \citenamefont {Yang},\ and\ \citenamefont {Jin}}]{Tang2018SA}%
  \BibitemOpen
  \bibfield  {author} {\bibinfo {author} {\bibfnamefont {H.}~\bibnamefont
  {Tang}}, \bibinfo {author} {\bibfnamefont {X.-F.}\ \bibnamefont {Lin}},
  \bibinfo {author} {\bibfnamefont {Z.}~\bibnamefont {Feng}}, \bibinfo {author}
  {\bibfnamefont {J.-Y.}\ \bibnamefont {Chen}}, \bibinfo {author}
  {\bibfnamefont {J.}~\bibnamefont {Gao}}, \bibinfo {author} {\bibfnamefont
  {K.}~\bibnamefont {Sun}}, \bibinfo {author} {\bibfnamefont {C.-Y.}\
  \bibnamefont {Wang}}, \bibinfo {author} {\bibfnamefont {P.-C.}\ \bibnamefont
  {Lai}}, \bibinfo {author} {\bibfnamefont {X.-Y.}\ \bibnamefont {Xu}},
  \bibinfo {author} {\bibfnamefont {Y.}~\bibnamefont {Wang}}, \bibinfo {author}
  {\bibfnamefont {L.-F.}\ \bibnamefont {Qiao}}, \bibinfo {author}
  {\bibfnamefont {A.-L.}\ \bibnamefont {Yang}}, \ and\ \bibinfo {author}
  {\bibfnamefont {X.-M.}\ \bibnamefont {Jin}},\ }\href {\doibase 10/gdjvss}
  {\bibfield  {journal} {\bibinfo  {journal} {Sci.\ Adv.}\ }\textbf {\bibinfo
  {volume} {4}},\ \bibinfo {pages} {eaat3174} (\bibinfo {year}
  {2018})}\BibitemShut {NoStop}%
\bibitem [{\citenamefont {{Venegas-Andraca}}(2012)}]{Venegas-Andraca2012QIP}%
  \BibitemOpen
  \bibfield  {author} {\bibinfo {author} {\bibfnamefont {S.~E.}\ \bibnamefont
  {{Venegas-Andraca}}},\ }\href {\doibase 10/b2gg} {\bibfield  {journal}
  {\bibinfo  {journal} {Quantum Inf.\ Process.}\ }\textbf {\bibinfo {volume}
  {11}},\ \bibinfo {pages} {1015} (\bibinfo {year} {2012})}\BibitemShut
  {NoStop}%
\bibitem [{\citenamefont {Farhi}\ and\ \citenamefont
  {Gutmann}(1998)}]{Farhi1998PRA}%
  \BibitemOpen
  \bibfield  {author} {\bibinfo {author} {\bibfnamefont {E.}~\bibnamefont
  {Farhi}}\ and\ \bibinfo {author} {\bibfnamefont {S.}~\bibnamefont
  {Gutmann}},\ }\href {\doibase 10/d9rf9j} {\bibfield  {journal} {\bibinfo
  {journal} {Phys.\ Rev.\ A}\ }\textbf {\bibinfo {volume} {58}},\ \bibinfo
  {pages} {915} (\bibinfo {year} {1998})}\BibitemShut {NoStop}%
\bibitem [{\citenamefont {Su}\ \emph {et~al.}(1979)\citenamefont {Su},
  \citenamefont {Schrieffer},\ and\ \citenamefont {Heeger}}]{Su1979PRL}%
  \BibitemOpen
  \bibfield  {author} {\bibinfo {author} {\bibfnamefont {W.~P.}\ \bibnamefont
  {Su}}, \bibinfo {author} {\bibfnamefont {J.~R.}\ \bibnamefont {Schrieffer}},
  \ and\ \bibinfo {author} {\bibfnamefont {A.~J.}\ \bibnamefont {Heeger}},\
  }\href {\doibase 10/fj76w9} {\bibfield  {journal} {\bibinfo  {journal}
  {Phys.\ Rev.\ Lett.}\ }\textbf {\bibinfo {volume} {42}},\ \bibinfo {pages}
  {1698} (\bibinfo {year} {1979})}\BibitemShut {NoStop}%
\bibitem [{\citenamefont {Schmitz}\ and\ \citenamefont
  {Schwalm}(2016)}]{Schmitz2016PLA}%
  \BibitemOpen
  \bibfield  {author} {\bibinfo {author} {\bibfnamefont {A.~T.}\ \bibnamefont
  {Schmitz}}\ and\ \bibinfo {author} {\bibfnamefont {W.~A.}\ \bibnamefont
  {Schwalm}},\ }\href {\doibase 10/f8b2xm} {\bibfield  {journal} {\bibinfo
  {journal} {Phys.\ Lett.\ A}\ }\textbf {\bibinfo {volume} {380}},\ \bibinfo
  {pages} {1125} (\bibinfo {year} {2016})}\BibitemShut {NoStop}%
\bibitem [{\citenamefont {Berry}(1984)}]{Berry1984PRSLA}%
  \BibitemOpen
  \bibfield  {author} {\bibinfo {author} {\bibfnamefont {M.~V.}\ \bibnamefont
  {Berry}},\ }\href {\doibase 10/cctx57} {\bibfield  {journal} {\bibinfo
  {journal} {Proc.\ Royal Soc.\ Lond.\ A}\ }\textbf {\bibinfo {volume} {392}},\
  \bibinfo {pages} {45} (\bibinfo {year} {1984})}\BibitemShut {NoStop}%
\bibitem [{\citenamefont {v.~Klitzing}\ \emph {et~al.}(1980)\citenamefont
  {v.~Klitzing}, \citenamefont {Dorda},\ and\ \citenamefont
  {Pepper}}]{Klitzing1980PRL}%
  \BibitemOpen
  \bibfield  {author} {\bibinfo {author} {\bibfnamefont {K.}~\bibnamefont
  {v.~Klitzing}}, \bibinfo {author} {\bibfnamefont {G.}~\bibnamefont {Dorda}},
  \ and\ \bibinfo {author} {\bibfnamefont {M.}~\bibnamefont {Pepper}},\ }\href
  {\doibase 10/bttjkh} {\bibfield  {journal} {\bibinfo  {journal} {Phys.\ Rev.\
  Lett.}\ }\textbf {\bibinfo {volume} {45}},\ \bibinfo {pages} {494} (\bibinfo
  {year} {1980})}\BibitemShut {NoStop}%
\bibitem [{\citenamefont {Chern}(1946)}]{Chern1946AM}%
  \BibitemOpen
  \bibfield  {author} {\bibinfo {author} {\bibfnamefont {S.-S.}\ \bibnamefont
  {Chern}},\ }\href {\doibase 10/bpx8rd} {\bibfield  {journal} {\bibinfo
  {journal} {Ann.\ Math.}\ }\textbf {\bibinfo {volume} {47}},\ \bibinfo {pages}
  {85} (\bibinfo {year} {1946})}\BibitemShut {NoStop}%
\bibitem [{\citenamefont {Chern}\ and\ \citenamefont
  {Simons}(1974)}]{Chern1974AM}%
  \BibitemOpen
  \bibfield  {author} {\bibinfo {author} {\bibfnamefont {S.-S.}\ \bibnamefont
  {Chern}}\ and\ \bibinfo {author} {\bibfnamefont {J.}~\bibnamefont {Simons}},\
  }\href {\doibase 10/fr886v} {\bibfield  {journal} {\bibinfo  {journal} {Ann.\
  Math.}\ }\textbf {\bibinfo {volume} {99}},\ \bibinfo {pages} {48} (\bibinfo
  {year} {1974})}\BibitemShut {NoStop}%
\bibitem [{\citenamefont {Cardano}\ \emph {et~al.}(2016)\citenamefont
  {Cardano}, \citenamefont {Maffei}, \citenamefont {Massa}, \citenamefont
  {Piccirillo}, \citenamefont {{de Lisio}}, \citenamefont {De~Filippis},
  \citenamefont {Cataudella}, \citenamefont {Santamato},\ and\ \citenamefont
  {Marrucci}}]{Cardano2016NC}%
  \BibitemOpen
  \bibfield  {author} {\bibinfo {author} {\bibfnamefont {F.}~\bibnamefont
  {Cardano}}, \bibinfo {author} {\bibfnamefont {M.}~\bibnamefont {Maffei}},
  \bibinfo {author} {\bibfnamefont {F.}~\bibnamefont {Massa}}, \bibinfo
  {author} {\bibfnamefont {B.}~\bibnamefont {Piccirillo}}, \bibinfo {author}
  {\bibfnamefont {C.}~\bibnamefont {{de Lisio}}}, \bibinfo {author}
  {\bibfnamefont {G.}~\bibnamefont {De~Filippis}}, \bibinfo {author}
  {\bibfnamefont {V.}~\bibnamefont {Cataudella}}, \bibinfo {author}
  {\bibfnamefont {E.}~\bibnamefont {Santamato}}, \ and\ \bibinfo {author}
  {\bibfnamefont {L.}~\bibnamefont {Marrucci}},\ }\href {\doibase
  10.1038/ncomms11439} {\bibfield  {journal} {\bibinfo  {journal} {Nat.\
  Commun.}\ }\textbf {\bibinfo {volume} {7}},\ \bibinfo {pages} {11439}
  (\bibinfo {year} {2016})}\BibitemShut {NoStop}%
\bibitem [{\citenamefont {Cardano}\ \emph {et~al.}(2017)\citenamefont
  {Cardano}, \citenamefont {D'Errico}, \citenamefont {Dauphin}, \citenamefont
  {Maffei}, \citenamefont {Piccirillo}, \citenamefont {{de Lisio}},
  \citenamefont {De~Filippis}, \citenamefont {Cataudella}, \citenamefont
  {Santamato}, \citenamefont {Marrucci}, \citenamefont {Lewenstein},\ and\
  \citenamefont {Massignan}}]{Cardano2017NC}%
  \BibitemOpen
  \bibfield  {author} {\bibinfo {author} {\bibfnamefont {F.}~\bibnamefont
  {Cardano}}, \bibinfo {author} {\bibfnamefont {A.}~\bibnamefont {D'Errico}},
  \bibinfo {author} {\bibfnamefont {A.}~\bibnamefont {Dauphin}}, \bibinfo
  {author} {\bibfnamefont {M.}~\bibnamefont {Maffei}}, \bibinfo {author}
  {\bibfnamefont {B.}~\bibnamefont {Piccirillo}}, \bibinfo {author}
  {\bibfnamefont {C.}~\bibnamefont {{de Lisio}}}, \bibinfo {author}
  {\bibfnamefont {G.}~\bibnamefont {De~Filippis}}, \bibinfo {author}
  {\bibfnamefont {V.}~\bibnamefont {Cataudella}}, \bibinfo {author}
  {\bibfnamefont {E.}~\bibnamefont {Santamato}}, \bibinfo {author}
  {\bibfnamefont {L.}~\bibnamefont {Marrucci}}, \bibinfo {author}
  {\bibfnamefont {M.}~\bibnamefont {Lewenstein}}, \ and\ \bibinfo {author}
  {\bibfnamefont {P.}~\bibnamefont {Massignan}},\ }\href {\doibase 10/f99bhf}
  {\bibfield  {journal} {\bibinfo  {journal} {Nat.\ Commun.}\ }\textbf
  {\bibinfo {volume} {8}},\ \bibinfo {pages} {15516} (\bibinfo {year}
  {2017})}\BibitemShut {NoStop}%
\bibitem [{\citenamefont {Wang}\ \emph
  {et~al.}(2019{\natexlab{a}})\citenamefont {Wang}, \citenamefont {Lu},
  \citenamefont {Mei}, \citenamefont {Gao}, \citenamefont {Li}, \citenamefont
  {Tang}, \citenamefont {Zhu}, \citenamefont {Jia},\ and\ \citenamefont
  {Jin}}]{Wang2019PRL}%
  \BibitemOpen
  \bibfield  {author} {\bibinfo {author} {\bibfnamefont {Y.}~\bibnamefont
  {Wang}}, \bibinfo {author} {\bibfnamefont {Y.-H.}\ \bibnamefont {Lu}},
  \bibinfo {author} {\bibfnamefont {F.}~\bibnamefont {Mei}}, \bibinfo {author}
  {\bibfnamefont {J.}~\bibnamefont {Gao}}, \bibinfo {author} {\bibfnamefont
  {Z.-M.}\ \bibnamefont {Li}}, \bibinfo {author} {\bibfnamefont
  {H.}~\bibnamefont {Tang}}, \bibinfo {author} {\bibfnamefont {S.-L.}\
  \bibnamefont {Zhu}}, \bibinfo {author} {\bibfnamefont {S.}~\bibnamefont
  {Jia}}, \ and\ \bibinfo {author} {\bibfnamefont {X.-M.}\ \bibnamefont
  {Jin}},\ }\href {\doibase 10/gf27nd} {\bibfield  {journal} {\bibinfo
  {journal} {Phys.\ Rev.\ Lett.}\ }\textbf {\bibinfo {volume} {122}},\ \bibinfo
  {pages} {193903} (\bibinfo {year} {2019}{\natexlab{a}})}\BibitemShut
  {NoStop}%
\bibitem [{\citenamefont {Groh}\ \emph {et~al.}(2016)\citenamefont {Groh},
  \citenamefont {Brakhane}, \citenamefont {Alt}, \citenamefont {Meschede},
  \citenamefont {Asb{\'o}th},\ and\ \citenamefont {Alberti}}]{Groh2016PRA}%
  \BibitemOpen
  \bibfield  {author} {\bibinfo {author} {\bibfnamefont {T.}~\bibnamefont
  {Groh}}, \bibinfo {author} {\bibfnamefont {S.}~\bibnamefont {Brakhane}},
  \bibinfo {author} {\bibfnamefont {W.}~\bibnamefont {Alt}}, \bibinfo {author}
  {\bibfnamefont {D.}~\bibnamefont {Meschede}}, \bibinfo {author}
  {\bibfnamefont {J.~K.}\ \bibnamefont {Asb{\'o}th}}, \ and\ \bibinfo {author}
  {\bibfnamefont {A.}~\bibnamefont {Alberti}},\ }\href {\doibase
  10.1103/PhysRevA.94.013620} {\bibfield  {journal} {\bibinfo  {journal}
  {Phys.\ Rev.\ A}\ }\textbf {\bibinfo {volume} {94}},\ \bibinfo {pages}
  {013620} (\bibinfo {year} {2016})}\BibitemShut {NoStop}%
\bibitem [{\citenamefont {Mugel}\ \emph {et~al.}(2016)\citenamefont {Mugel},
  \citenamefont {Celi}, \citenamefont {Massignan}, \citenamefont {Asb{\'o}th},
  \citenamefont {Lewenstein},\ and\ \citenamefont {Lobo}}]{Mugel2016PRA}%
  \BibitemOpen
  \bibfield  {author} {\bibinfo {author} {\bibfnamefont {S.}~\bibnamefont
  {Mugel}}, \bibinfo {author} {\bibfnamefont {A.}~\bibnamefont {Celi}},
  \bibinfo {author} {\bibfnamefont {P.}~\bibnamefont {Massignan}}, \bibinfo
  {author} {\bibfnamefont {J.~K.}\ \bibnamefont {Asb{\'o}th}}, \bibinfo
  {author} {\bibfnamefont {M.}~\bibnamefont {Lewenstein}}, \ and\ \bibinfo
  {author} {\bibfnamefont {C.}~\bibnamefont {Lobo}},\ }\href {\doibase
  10/gf5ggn} {\bibfield  {journal} {\bibinfo  {journal} {Phys. Rev. A}\
  }\textbf {\bibinfo {volume} {94}},\ \bibinfo {pages} {023631} (\bibinfo
  {year} {2016})}\BibitemShut {NoStop}%
\bibitem [{\citenamefont {Nitsche}\ \emph {et~al.}(2019)\citenamefont
  {Nitsche}, \citenamefont {Geib}, \citenamefont {Stahl}, \citenamefont {Lorz},
  \citenamefont {Cedzich}, \citenamefont {Barkhofen}, \citenamefont {Werner},\
  and\ \citenamefont {Silberhorn}}]{Nitsche2019NJP}%
  \BibitemOpen
  \bibfield  {author} {\bibinfo {author} {\bibfnamefont {T.}~\bibnamefont
  {Nitsche}}, \bibinfo {author} {\bibfnamefont {T.}~\bibnamefont {Geib}},
  \bibinfo {author} {\bibfnamefont {C.}~\bibnamefont {Stahl}}, \bibinfo
  {author} {\bibfnamefont {L.}~\bibnamefont {Lorz}}, \bibinfo {author}
  {\bibfnamefont {C.}~\bibnamefont {Cedzich}}, \bibinfo {author} {\bibfnamefont
  {S.}~\bibnamefont {Barkhofen}}, \bibinfo {author} {\bibfnamefont {R.~F.}\
  \bibnamefont {Werner}}, \ and\ \bibinfo {author} {\bibfnamefont
  {C.}~\bibnamefont {Silberhorn}},\ }\href {\doibase 10/gf2wn2} {\bibfield
  {journal} {\bibinfo  {journal} {New J.\ Phys.}\ }\textbf {\bibinfo {volume}
  {21}},\ \bibinfo {pages} {043031} (\bibinfo {year} {2019})}\BibitemShut
  {NoStop}%
\bibitem [{\citenamefont {Zhang}\ \emph
  {et~al.}(2017{\natexlab{b}})\citenamefont {Zhang}, \citenamefont {Goyal},
  \citenamefont {Simon},\ and\ \citenamefont {Sanders}}]{Zhang2017PRA}%
  \BibitemOpen
  \bibfield  {author} {\bibinfo {author} {\bibfnamefont {W.-W.}\ \bibnamefont
  {Zhang}}, \bibinfo {author} {\bibfnamefont {S.~K.}\ \bibnamefont {Goyal}},
  \bibinfo {author} {\bibfnamefont {C.}~\bibnamefont {Simon}}, \ and\ \bibinfo
  {author} {\bibfnamefont {B.~C.}\ \bibnamefont {Sanders}},\ }\href {\doibase
  10/gd7fdz} {\bibfield  {journal} {\bibinfo  {journal} {Phys.\ Rev.\ A}\
  }\textbf {\bibinfo {volume} {95}},\ \bibinfo {pages} {052351} (\bibinfo
  {year} {2017}{\natexlab{b}})}\BibitemShut {NoStop}%
\bibitem [{\citenamefont {Yao}\ \emph {et~al.}(2017)\citenamefont {Yao},
  \citenamefont {Yan},\ and\ \citenamefont {Wang}}]{Yao2017PRB}%
  \BibitemOpen
  \bibfield  {author} {\bibinfo {author} {\bibfnamefont {S.}~\bibnamefont
  {Yao}}, \bibinfo {author} {\bibfnamefont {Z.}~\bibnamefont {Yan}}, \ and\
  \bibinfo {author} {\bibfnamefont {Z.}~\bibnamefont {Wang}},\ }\href {\doibase
  10/gf24qn} {\bibfield  {journal} {\bibinfo  {journal} {Phys.\ Rev.\ B}\
  }\textbf {\bibinfo {volume} {96}},\ \bibinfo {pages} {195303} (\bibinfo
  {year} {2017})}\BibitemShut {NoStop}%
\bibitem [{\citenamefont {Tarasinski}\ \emph {et~al.}(2014)\citenamefont
  {Tarasinski}, \citenamefont {Asb{\'o}th},\ and\ \citenamefont
  {Dahlhaus}}]{Tarasinski2014PRA}%
  \BibitemOpen
  \bibfield  {author} {\bibinfo {author} {\bibfnamefont {B.}~\bibnamefont
  {Tarasinski}}, \bibinfo {author} {\bibfnamefont {J.~K.}\ \bibnamefont
  {Asb{\'o}th}}, \ and\ \bibinfo {author} {\bibfnamefont {J.~P.}\ \bibnamefont
  {Dahlhaus}},\ }\href {\doibase 10/gf2ngv} {\bibfield  {journal} {\bibinfo
  {journal} {Phys.\ Rev.\ A}\ }\textbf {\bibinfo {volume} {89}},\ \bibinfo
  {pages} {042327} (\bibinfo {year} {2014})}\BibitemShut {NoStop}%
\bibitem [{\citenamefont {Barkhofen}\ \emph {et~al.}(2017)\citenamefont
  {Barkhofen}, \citenamefont {Nitsche}, \citenamefont {Elster}, \citenamefont
  {Lorz}, \citenamefont {G{\'a}bris}, \citenamefont {Jex},\ and\ \citenamefont
  {Silberhorn}}]{Barkhofen2017PRA}%
  \BibitemOpen
  \bibfield  {author} {\bibinfo {author} {\bibfnamefont {S.}~\bibnamefont
  {Barkhofen}}, \bibinfo {author} {\bibfnamefont {T.}~\bibnamefont {Nitsche}},
  \bibinfo {author} {\bibfnamefont {F.}~\bibnamefont {Elster}}, \bibinfo
  {author} {\bibfnamefont {L.}~\bibnamefont {Lorz}}, \bibinfo {author}
  {\bibfnamefont {A.}~\bibnamefont {G{\'a}bris}}, \bibinfo {author}
  {\bibfnamefont {I.}~\bibnamefont {Jex}}, \ and\ \bibinfo {author}
  {\bibfnamefont {C.}~\bibnamefont {Silberhorn}},\ }\href {\doibase 10/gf5grk}
  {\bibfield  {journal} {\bibinfo  {journal} {Phys.\ Rev.\ A}\ }\textbf
  {\bibinfo {volume} {96}},\ \bibinfo {pages} {033846} (\bibinfo {year}
  {2017})}\BibitemShut {NoStop}%
\bibitem [{\citenamefont {Asb\'oth}\ and\ \citenamefont
  {Obuse}(2013)}]{Asboth2013PRB}%
  \BibitemOpen
  \bibfield  {author} {\bibinfo {author} {\bibfnamefont {J.~K.}\ \bibnamefont
  {Asb\'oth}}\ and\ \bibinfo {author} {\bibfnamefont {H.}~\bibnamefont
  {Obuse}},\ }\href {\doibase 10/gfvqnb} {\bibfield  {journal} {\bibinfo
  {journal} {Phys.\ Rev.\ B}\ }\textbf {\bibinfo {volume} {88}},\ \bibinfo
  {pages} {121406} (\bibinfo {year} {2013})}\BibitemShut {NoStop}%
\bibitem [{\citenamefont {Asb{\'o}th}\ \emph {et~al.}(2014)\citenamefont
  {Asb{\'o}th}, \citenamefont {Tarasinski},\ and\ \citenamefont
  {Delplace}}]{Asboth2014PRB}%
  \BibitemOpen
  \bibfield  {author} {\bibinfo {author} {\bibfnamefont {J.~K.}\ \bibnamefont
  {Asb{\'o}th}}, \bibinfo {author} {\bibfnamefont {B.}~\bibnamefont
  {Tarasinski}}, \ and\ \bibinfo {author} {\bibfnamefont {P.}~\bibnamefont
  {Delplace}},\ }\href {\doibase 10/gfzr4b} {\bibfield  {journal} {\bibinfo
  {journal} {Phys.\ Rev.\ B}\ }\textbf {\bibinfo {volume} {90}},\ \bibinfo
  {pages} {125143} (\bibinfo {year} {2014})}\BibitemShut {NoStop}%
\bibitem [{\citenamefont {Obuse}\ \emph {et~al.}(2015)\citenamefont {Obuse},
  \citenamefont {Asb{\'o}th}, \citenamefont {Nishimura},\ and\ \citenamefont
  {Kawakami}}]{Obuse2015PRB}%
  \BibitemOpen
  \bibfield  {author} {\bibinfo {author} {\bibfnamefont {H.}~\bibnamefont
  {Obuse}}, \bibinfo {author} {\bibfnamefont {J.~K.}\ \bibnamefont
  {Asb{\'o}th}}, \bibinfo {author} {\bibfnamefont {Y.}~\bibnamefont
  {Nishimura}}, \ and\ \bibinfo {author} {\bibfnamefont {N.}~\bibnamefont
  {Kawakami}},\ }\href {\doibase 10/gf5gwn} {\bibfield  {journal} {\bibinfo
  {journal} {Phys.\ Rev.\ B}\ }\textbf {\bibinfo {volume} {92}},\ \bibinfo
  {pages} {045424} (\bibinfo {year} {2015})}\BibitemShut {NoStop}%
\bibitem [{\citenamefont {Cedzich}\ \emph {et~al.}(2016)\citenamefont
  {Cedzich}, \citenamefont {Gr{\"u}nbaum}, \citenamefont {Stahl}, \citenamefont
  {Vel{\'a}zquez}, \citenamefont {Werner},\ and\ \citenamefont
  {Werner}}]{Cedzich2016JPA}%
  \BibitemOpen
  \bibfield  {author} {\bibinfo {author} {\bibfnamefont {C.}~\bibnamefont
  {Cedzich}}, \bibinfo {author} {\bibfnamefont {F.~A.}\ \bibnamefont
  {Gr{\"u}nbaum}}, \bibinfo {author} {\bibfnamefont {C.}~\bibnamefont {Stahl}},
  \bibinfo {author} {\bibfnamefont {L.}~\bibnamefont {Vel{\'a}zquez}}, \bibinfo
  {author} {\bibfnamefont {A.~H.}\ \bibnamefont {Werner}}, \ and\ \bibinfo
  {author} {\bibfnamefont {R.~F.}\ \bibnamefont {Werner}},\ }\href {\doibase
  10/gf3bhr} {\bibfield  {journal} {\bibinfo  {journal} {J.\ Phys.\ A}\
  }\textbf {\bibinfo {volume} {49}},\ \bibinfo {pages} {21LT01} (\bibinfo
  {year} {2016})}\BibitemShut {NoStop}%
\bibitem [{\citenamefont {Cedzich}\ \emph
  {et~al.}(2018{\natexlab{a}})\citenamefont {Cedzich}, \citenamefont {Geib},
  \citenamefont {Gr{\"u}nbaum}, \citenamefont {Stahl}, \citenamefont
  {Vel{\'a}zquez}, \citenamefont {Werner},\ and\ \citenamefont
  {Werner}}]{Cedzich2018AHP}%
  \BibitemOpen
  \bibfield  {author} {\bibinfo {author} {\bibfnamefont {C.}~\bibnamefont
  {Cedzich}}, \bibinfo {author} {\bibfnamefont {T.}~\bibnamefont {Geib}},
  \bibinfo {author} {\bibfnamefont {F.~A.}\ \bibnamefont {Gr{\"u}nbaum}},
  \bibinfo {author} {\bibfnamefont {C.}~\bibnamefont {Stahl}}, \bibinfo
  {author} {\bibfnamefont {L.}~\bibnamefont {Vel{\'a}zquez}}, \bibinfo {author}
  {\bibfnamefont {A.~H.}\ \bibnamefont {Werner}}, \ and\ \bibinfo {author}
  {\bibfnamefont {R.~F.}\ \bibnamefont {Werner}},\ }\href {\doibase 10/gcw4fs}
  {\bibfield  {journal} {\bibinfo  {journal} {Ann.\ Henri Poincar{\'e}}\
  }\textbf {\bibinfo {volume} {19}},\ \bibinfo {pages} {325} (\bibinfo {year}
  {2018}{\natexlab{a}})}\BibitemShut {NoStop}%
\bibitem [{\citenamefont {Cedzich}\ \emph
  {et~al.}(2018{\natexlab{b}})\citenamefont {Cedzich}, \citenamefont {Geib},
  \citenamefont {Stahl}, \citenamefont {Vel{\'a}zquez}, \citenamefont
  {Werner},\ and\ \citenamefont {Werner}}]{Cedzich2018Quantum}%
  \BibitemOpen
  \bibfield  {author} {\bibinfo {author} {\bibfnamefont {C.}~\bibnamefont
  {Cedzich}}, \bibinfo {author} {\bibfnamefont {T.}~\bibnamefont {Geib}},
  \bibinfo {author} {\bibfnamefont {C.}~\bibnamefont {Stahl}}, \bibinfo
  {author} {\bibfnamefont {L.}~\bibnamefont {Vel{\'a}zquez}}, \bibinfo {author}
  {\bibfnamefont {A.~H.}\ \bibnamefont {Werner}}, \ and\ \bibinfo {author}
  {\bibfnamefont {R.~F.}\ \bibnamefont {Werner}},\ }\href {\doibase 10/gfz8r8}
  {\bibfield  {journal} {\bibinfo  {journal} {Quantum}\ }\textbf {\bibinfo
  {volume} {2}},\ \bibinfo {pages} {95} (\bibinfo {year}
  {2018}{\natexlab{b}})}\BibitemShut {NoStop}%
\bibitem [{\citenamefont {Cedzich}\ \emph {et~al.}(2019)\citenamefont
  {Cedzich}, \citenamefont {Geib}, \citenamefont {Gr{\"u}nbaum}, \citenamefont
  {Vel{\'a}zquez}, \citenamefont {Werner},\ and\ \citenamefont
  {Werner}}]{Cedzich2019arXiv}%
  \BibitemOpen
  \bibfield  {author} {\bibinfo {author} {\bibfnamefont {C.}~\bibnamefont
  {Cedzich}}, \bibinfo {author} {\bibfnamefont {T.}~\bibnamefont {Geib}},
  \bibinfo {author} {\bibfnamefont {F.~A.}\ \bibnamefont {Gr{\"u}nbaum}},
  \bibinfo {author} {\bibfnamefont {L.}~\bibnamefont {Vel{\'a}zquez}}, \bibinfo
  {author} {\bibfnamefont {A.~H.}\ \bibnamefont {Werner}}, \ and\ \bibinfo
  {author} {\bibfnamefont {R.~F.}\ \bibnamefont {Werner}},\ }\href@noop {}
  {\enquote {\bibinfo {title} {Quantum walks: {Schur} functions meet symmetry
  protected topological phases},}\ } (\bibinfo {year} {2019}),\ \Eprint
  {http://arxiv.org/abs/1903.07494} {arXiv:1903.07494 [math-ph]} \BibitemShut
  {NoStop}%
\bibitem [{\citenamefont {Asb{\'o}th}\ and\ \citenamefont
  {Edge}(2015)}]{Asboth2015PRA}%
  \BibitemOpen
  \bibfield  {author} {\bibinfo {author} {\bibfnamefont {J.~K.}\ \bibnamefont
  {Asb{\'o}th}}\ and\ \bibinfo {author} {\bibfnamefont {J.~M.}\ \bibnamefont
  {Edge}},\ }\href {\doibase 10/f3nj9n} {\bibfield  {journal} {\bibinfo
  {journal} {Phys.\ Rev.\ A}\ }\textbf {\bibinfo {volume} {91}},\ \bibinfo
  {pages} {022324} (\bibinfo {year} {2015})}\BibitemShut {NoStop}%
\bibitem [{\citenamefont {Wang}\ \emph {et~al.}(2018)\citenamefont {Wang},
  \citenamefont {Chen},\ and\ \citenamefont {Zhang}}]{Wang2018PRLb}%
  \BibitemOpen
  \bibfield  {author} {\bibinfo {author} {\bibfnamefont {B.}~\bibnamefont
  {Wang}}, \bibinfo {author} {\bibfnamefont {T.}~\bibnamefont {Chen}}, \ and\
  \bibinfo {author} {\bibfnamefont {X.}~\bibnamefont {Zhang}},\ }\href
  {\doibase 10/gd55m4} {\bibfield  {journal} {\bibinfo  {journal} {Phys.\ Rev.\
  Lett.}\ }\textbf {\bibinfo {volume} {121}},\ \bibinfo {pages} {100501}
  (\bibinfo {year} {2018})}\BibitemShut {NoStop}%
\bibitem [{\citenamefont {Xu}\ \emph {et~al.}(2018)\citenamefont {Xu},
  \citenamefont {Wang}, \citenamefont {Pan}, \citenamefont {Sun}, \citenamefont
  {Xu}, \citenamefont {Chen}, \citenamefont {Tang}, \citenamefont {Gong},
  \citenamefont {Han}, \citenamefont {Li},\ and\ \citenamefont
  {Guo}}]{Xu2018PRL}%
  \BibitemOpen
  \bibfield  {author} {\bibinfo {author} {\bibfnamefont {X.-Y.}\ \bibnamefont
  {Xu}}, \bibinfo {author} {\bibfnamefont {Q.-Q.}\ \bibnamefont {Wang}},
  \bibinfo {author} {\bibfnamefont {W.-W.}\ \bibnamefont {Pan}}, \bibinfo
  {author} {\bibfnamefont {K.}~\bibnamefont {Sun}}, \bibinfo {author}
  {\bibfnamefont {J.-S.}\ \bibnamefont {Xu}}, \bibinfo {author} {\bibfnamefont
  {G.}~\bibnamefont {Chen}}, \bibinfo {author} {\bibfnamefont {J.-S.}\
  \bibnamefont {Tang}}, \bibinfo {author} {\bibfnamefont {M.}~\bibnamefont
  {Gong}}, \bibinfo {author} {\bibfnamefont {Y.-J.}\ \bibnamefont {Han}},
  \bibinfo {author} {\bibfnamefont {C.-F.}\ \bibnamefont {Li}}, \ and\ \bibinfo
  {author} {\bibfnamefont {G.-C.}\ \bibnamefont {Guo}},\ }\href {\doibase
  10/gdtgs5} {\bibfield  {journal} {\bibinfo  {journal} {Phys. Rev. Lett.}\
  }\textbf {\bibinfo {volume} {120}},\ \bibinfo {pages} {260501} (\bibinfo
  {year} {2018})}\BibitemShut {NoStop}%
\bibitem [{\citenamefont {Ramasesh}\ \emph {et~al.}(2017)\citenamefont
  {Ramasesh}, \citenamefont {Flurin}, \citenamefont {Rudner}, \citenamefont
  {Siddiqi},\ and\ \citenamefont {Yao}}]{Ramasesh2017PRL}%
  \BibitemOpen
  \bibfield  {author} {\bibinfo {author} {\bibfnamefont {V.~V.}\ \bibnamefont
  {Ramasesh}}, \bibinfo {author} {\bibfnamefont {E.}~\bibnamefont {Flurin}},
  \bibinfo {author} {\bibfnamefont {M.}~\bibnamefont {Rudner}}, \bibinfo
  {author} {\bibfnamefont {I.}~\bibnamefont {Siddiqi}}, \ and\ \bibinfo
  {author} {\bibfnamefont {N.~Y.}\ \bibnamefont {Yao}},\ }\href {\doibase
  10.1103/PhysRevLett.118.130501} {\bibfield  {journal} {\bibinfo  {journal}
  {Phys.\ Rev.\ Lett.}\ }\textbf {\bibinfo {volume} {118}},\ \bibinfo {pages}
  {130501} (\bibinfo {year} {2017})}\BibitemShut {NoStop}%
\bibitem [{\citenamefont {{Blanco-Redondo}}\ \emph {et~al.}(2016)\citenamefont
  {{Blanco-Redondo}}, \citenamefont {Andonegui}, \citenamefont {Collins},
  \citenamefont {Harari}, \citenamefont {Lumer}, \citenamefont {Rechtsman},
  \citenamefont {Eggleton},\ and\ \citenamefont
  {Segev}}]{Blanco-Redondo2016PRL}%
  \BibitemOpen
  \bibfield  {author} {\bibinfo {author} {\bibfnamefont {A.}~\bibnamefont
  {{Blanco-Redondo}}}, \bibinfo {author} {\bibfnamefont {I.}~\bibnamefont
  {Andonegui}}, \bibinfo {author} {\bibfnamefont {M.~J.}\ \bibnamefont
  {Collins}}, \bibinfo {author} {\bibfnamefont {G.}~\bibnamefont {Harari}},
  \bibinfo {author} {\bibfnamefont {Y.}~\bibnamefont {Lumer}}, \bibinfo
  {author} {\bibfnamefont {M.~C.}\ \bibnamefont {Rechtsman}}, \bibinfo {author}
  {\bibfnamefont {B.~J.}\ \bibnamefont {Eggleton}}, \ and\ \bibinfo {author}
  {\bibfnamefont {M.}~\bibnamefont {Segev}},\ }\href {\doibase
  10.1103/PhysRevLett.116.163901} {\bibfield  {journal} {\bibinfo  {journal}
  {Phys.\ Rev.\ Lett.}\ }\textbf {\bibinfo {volume} {116}},\ \bibinfo {pages}
  {163901} (\bibinfo {year} {2016})}\BibitemShut {NoStop}%
\bibitem [{\citenamefont {Zhang}\ \emph {et~al.}(2018)\citenamefont {Zhang},
  \citenamefont {Zhang}, \citenamefont {Niu},\ and\ \citenamefont
  {Liu}}]{Zhang2018SB}%
  \BibitemOpen
  \bibfield  {author} {\bibinfo {author} {\bibfnamefont {L.}~\bibnamefont
  {Zhang}}, \bibinfo {author} {\bibfnamefont {L.}~\bibnamefont {Zhang}},
  \bibinfo {author} {\bibfnamefont {S.}~\bibnamefont {Niu}}, \ and\ \bibinfo
  {author} {\bibfnamefont {X.-J.}\ \bibnamefont {Liu}},\ }\href {\doibase
  10.1016/j.scib.2018.09.018} {\bibfield  {journal} {\bibinfo  {journal} {Sci.\
  Bull.}\ }\textbf {\bibinfo {volume} {63}},\ \bibinfo {pages} {1385} (\bibinfo
  {year} {2018})}\BibitemShut {NoStop}%
\bibitem [{\citenamefont {Bender}\ and\ \citenamefont
  {Boettcher}(1998)}]{Bender1998PRL}%
  \BibitemOpen
  \bibfield  {author} {\bibinfo {author} {\bibfnamefont {C.~M.}\ \bibnamefont
  {Bender}}\ and\ \bibinfo {author} {\bibfnamefont {S.}~\bibnamefont
  {Boettcher}},\ }\href {\doibase 10/d4jsb5} {\bibfield  {journal} {\bibinfo
  {journal} {Phys.\ Rev.\ Lett.}\ }\textbf {\bibinfo {volume} {80}},\ \bibinfo
  {pages} {5243} (\bibinfo {year} {1998})}\BibitemShut {NoStop}%
\bibitem [{\citenamefont {Mochizuki}\ \emph {et~al.}(2016)\citenamefont
  {Mochizuki}, \citenamefont {Kim},\ and\ \citenamefont
  {Obuse}}]{Mochizuki2016PRA}%
  \BibitemOpen
  \bibfield  {author} {\bibinfo {author} {\bibfnamefont {K.}~\bibnamefont
  {Mochizuki}}, \bibinfo {author} {\bibfnamefont {D.}~\bibnamefont {Kim}}, \
  and\ \bibinfo {author} {\bibfnamefont {H.}~\bibnamefont {Obuse}},\ }\href
  {\doibase 10/gfvvdt} {\bibfield  {journal} {\bibinfo  {journal} {Phys.\ Rev.\
  A}\ }\textbf {\bibinfo {volume} {93}},\ \bibinfo {pages} {062116} (\bibinfo
  {year} {2016})}\BibitemShut {NoStop}%
\bibitem [{\citenamefont {Wang}\ \emph
  {et~al.}(2019{\natexlab{b}})\citenamefont {Wang}, \citenamefont {Qiu},
  \citenamefont {Xiao}, \citenamefont {Zhan}, \citenamefont {Bian},
  \citenamefont {Sanders}, \citenamefont {Yi},\ and\ \citenamefont
  {Xue}}]{Wang2019NC}%
  \BibitemOpen
  \bibfield  {author} {\bibinfo {author} {\bibfnamefont {K.}~\bibnamefont
  {Wang}}, \bibinfo {author} {\bibfnamefont {X.}~\bibnamefont {Qiu}}, \bibinfo
  {author} {\bibfnamefont {L.}~\bibnamefont {Xiao}}, \bibinfo {author}
  {\bibfnamefont {X.}~\bibnamefont {Zhan}}, \bibinfo {author} {\bibfnamefont
  {Z.}~\bibnamefont {Bian}}, \bibinfo {author} {\bibfnamefont {B.~C.}\
  \bibnamefont {Sanders}}, \bibinfo {author} {\bibfnamefont {W.}~\bibnamefont
  {Yi}}, \ and\ \bibinfo {author} {\bibfnamefont {P.}~\bibnamefont {Xue}},\
  }\href {\doibase 10/gf276r} {\bibfield  {journal} {\bibinfo  {journal} {Nat.\
  Commun.}\ }\textbf {\bibinfo {volume} {10}},\ \bibinfo {pages} {2293}
  (\bibinfo {year} {2019}{\natexlab{b}})}\BibitemShut {NoStop}%
\bibitem [{\citenamefont {Ming}\ \emph {et~al.}(2018)\citenamefont {Ming},
  \citenamefont {Lin}, \citenamefont {Bartlett},\ and\ \citenamefont
  {Zhang}}]{Ming2018arXiv}%
  \BibitemOpen
  \bibfield  {author} {\bibinfo {author} {\bibfnamefont {Y.}~\bibnamefont
  {Ming}}, \bibinfo {author} {\bibfnamefont {C.-T.}\ \bibnamefont {Lin}},
  \bibinfo {author} {\bibfnamefont {S.~D.}\ \bibnamefont {Bartlett}}, \ and\
  \bibinfo {author} {\bibfnamefont {W.-W.}\ \bibnamefont {Zhang}},\ }\href@noop
  {} {\enquote {\bibinfo {title} {Quantum topology identification with deep
  neural networks and quantum walks},}\ } (\bibinfo {year} {2018}),\ \Eprint
  {http://arxiv.org/abs/1811.12630} {arXiv:1811.12630 [quant-ph]} \BibitemShut
  {NoStop}%
\bibitem [{\citenamefont {Rem}\ \emph {et~al.}(2019)\citenamefont {Rem},
  \citenamefont {K{\"a}ming}, \citenamefont {Tarnowski}, \citenamefont
  {Asteria}, \citenamefont {Fl{\"a}schner}, \citenamefont {Becker},
  \citenamefont {Sengstock},\ and\ \citenamefont {Weitenberg}}]{Rem2019NP}%
  \BibitemOpen
  \bibfield  {author} {\bibinfo {author} {\bibfnamefont {B.~S.}\ \bibnamefont
  {Rem}}, \bibinfo {author} {\bibfnamefont {N.}~\bibnamefont {K{\"a}ming}},
  \bibinfo {author} {\bibfnamefont {M.}~\bibnamefont {Tarnowski}}, \bibinfo
  {author} {\bibfnamefont {L.}~\bibnamefont {Asteria}}, \bibinfo {author}
  {\bibfnamefont {N.}~\bibnamefont {Fl{\"a}schner}}, \bibinfo {author}
  {\bibfnamefont {C.}~\bibnamefont {Becker}}, \bibinfo {author} {\bibfnamefont
  {K.}~\bibnamefont {Sengstock}}, \ and\ \bibinfo {author} {\bibfnamefont
  {C.}~\bibnamefont {Weitenberg}},\ }\href {\doibase
  doi:10.1038/s41567-019-0554-0} {\bibfield  {journal} {\bibinfo  {journal}
  {Nat.\ Phys.}\ } (\bibinfo {year} {2019}),\ doi:10.1038/s41567-019-0554-0},\
  \bibinfo {note} {advance online publication}\BibitemShut {NoStop}%
\end{thebibliography}%
\end{document}